\documentclass[%
superscriptaddress,
amsmath,amssymb,
aps,
pra,
]{revtex4-2}

\usepackage{graphicx}
\usepackage{dcolumn}
\usepackage{bm}
\usepackage[caption=false]{subfig} 
\usepackage[colorlinks=true]{hyperref} 
\hypersetup{colorlinks,
	citecolor=blue
}
\usepackage[dvipsnames]{xcolor}
\definecolor{blue2}{rgb}{0.07, 0.04, 0.56}

\usepackage{booktabs} 
\usepackage{array}
\usepackage{tikz}
\usepackage{cancel}
\usetikzlibrary{calc,shapes.misc,shapes.geometric}
\newcommand\Mark[2][]{%
\tikz[baseline=(a.base)]{
\node[inner sep=0pt,outer sep=0pt](a){\phantom{#2}};  
\node[draw,black,thick,inner sep=2pt, rectangle,text=black,overlay,#1]  {#2};%
}
}

\newcommand{\eq}[1]{Eq.~\eqref{#1}}

\begin{document}

\title{Theoretical description of semi-inclusive T2K, Miner$\nu$A and MicroBooNE
neutrino-nucleus data in the relativistic plane wave impulse approximation}

\author{J. M. Franco-Patino}
\affiliation{
	Departamento de F\'isica at\'omica, molecular y nuclear, Universidad de Sevilla, 41080 Sevilla, Spain
}
\author{M. B. Barbaro}
\affiliation{
	Dipartimento di Fisica, Universit\`a di Torino and INFN, Sezione di Torino, Via P. Giuria 1, 10125 Torino, Italy
}
\affiliation{Universit\'e Paris-Saclay, CNRS/IN2P3, IJCLab, 91405 Orsay, France}
\author{J. A. Caballero}
\affiliation{
	Departamento de F\'isica at\'omica, molecular y nuclear, Universidad de Sevilla, 41080 Sevilla, Spain
}
\affiliation{
	Instituto de F\'isica Te\'orica y Computacional Carlos I, Granada 18071, Spain
}

\author{G. D. Megias}
\affiliation{
	Research Center for Cosmic Neutrinos, Institute for Cosmic Ray Research, University of Tokyo, Kashiwa, Chiba 277-8582, Japan
}
\affiliation{
	Departamento de F\'isica at\'omica, molecular y nuclear, Universidad de Sevilla, 41080 Sevilla, Spain
}
\date{\today}

\begin{abstract}
We present the results of semi-inclusive neutrino-nucleus cross sections within the plane wave impulse approximation (PWIA) for three nuclear models: relativistic Fermi gas (RFG), independent-particle shell model (IPSM) and natural orbital shell model (NO) in comparison with the available CC0$\pi$ measurements from the T2K, MINER$\nu$A and MicroBooNE collaborations where
a muon and at least one proton are detected in the final state.
Results are presented as a function of the momenta and angles of the final particles, as well as in terms of the imbalances between proton and muon kinematics. The analysis reveals that contributions beyond PWIA are crucial to explain the experimental measurements and that the study of correlations between final-state proton and muon kinematics can provide valuable
information on relevant nuclear effects such as the Fermi motion and final state interactions.   
\end{abstract}

\maketitle


\section{\label{sec:0}Introduction}
Neutrino oscillation is a valuable tool that can be used for extracting neutrino mixing angles, mass-squared
differences and the CP-symmetry violation phase, as well as for
looking for hints of new physics beyond the standard model in the
electroweak sector \cite{review,nature1}.

The oscillation probability is a function of the neutrino propagation distance and of its energy. In accelerator-based oscillation experiments, the neutrino propagation distance
is well defined. However, as these experiments do not use monochromatic neutrino beams, the accuracy to which they can extract neutrino oscillation parameters depends on their ability to determine the energy of the incoming neutrinos. This
relies on a proper understanding of the scattering of neutrinos with nucleons in the target and of the nuclear-medium effects involved, which are among the most relevant limiting factors for oscillation measurements.
Experimentally, the energy of incoming neutrinos is reconstructed from the particles generated after 
their interaction with the nuclear target. Therefore, in order to reduce the associated systematic uncertainties, it is essential to get a deep knowledge of neutrino interactions with nuclei, such as $^{12}$C, $^{16}$O or $^{40}$Ar, that are 
commonly used as targets. In particular, ongoing oscillation experiments are mainly focused on charged-current (CC) neutrino-nucleon quasielastic (QE) scattering interactions, where the neutrino removes a single nucleon from the nucleus without producing any additional particles. These reactions can be reasonably well approximated as two-body interactions, and their experimental signature of an identifiable lepton is relatively straightforward to measure. However, other processes such as multinucleon excitations - mainly the excitation of two-particle-two-hole (2p2h) states - of the nucleus, where only a charged lepton is emitted, can mimic a CCQE event in the
detectors, thus affecting the determination of the neutrino energy. As a result, the CCQE reaction is not directly accessible and what is experimentally measured are the so-called CC0$\pi$ events, $i.e.$, CC interactions with a charged lepton and no pions detected in the final state. 
Multi-nucleon excitations together with NN correlations and other effects related to the propagation of the knock-out nucleon through the nuclear medium must be included in the data analysis. 
CC0$\pi$ observables have been widely measured~\cite{AguilarArevalo:2010zc,AguilarArevalo:2013hm,Lyubushkin:2008pe,Minerva1,Minerva2,Minervanue,T2Kincl,T2Kinclelectron,T2Kcc0pi,Argoneutincl2} in terms of final lepton kinematics, yet an accurate identification of the different nuclear processes has been found difficult. This is because the lepton final state kinematics is largely affected by the nuclear dynamics and by the intrinsic dynamics of all particles involved.

In this context, one way to improve significantly the reconstruction of the neutrino energy together with the experimental systematics is the analysis of more exclusive processes where, in addition to the
final lepton as in purely inclusive measurements,
other particles are detected. This is, e.g., the case of semi-inclusive events in which the lepton is detected in coincidence with one hadron in the final state. Accordingly, the T2K, MINER$\nu$A and MicroBooNE collaborations are now focusing on the combined measurements of the charged lepton with one or more particles in coincidence~\cite{PhysRevD.98.032003,PhysRevLett.121.022504,PhysRevD.101.092001,microbooneA,microbooneB}. Also, forthcoming projects and experiments such as SK-Gd~\cite{Simpson:2019xwo}, HyperKamiokande~\cite{protocollaboration2018hyperkamiokande} and DUNE~\cite{dunecollaboration2016longbaseline} are making important efforts to improve the detection and identification capabilities of final-state hadrons. Thus, it is crucial to have realistic theoretical
nuclear models for the description of semi-inclusive observables, whose formalism is more complex 
but that make possible to analyze nuclear effects not accessible 
with inclusive processes. 

In Ref.~\cite{PhysRevC.102.064626} we presented a study of semi-inclusive CC neutrino-nucleus reactions in the plane wave impulse approximation (PWIA) for several nuclear models. In this work we apply this analysis to compare our predictions with the available semi-inclusive experimental data where a muon and at least one proton are detected in the final state. 
Although being aware of the oversimplified description of the reaction provided by PWIA, the present study is meant to be a first step towards a more complete modelling and a useful benchmark for more sophisticated calculations.
Results are compared with recent data from the T2K \cite{PhysRevD.98.032003}, MINER$\nu$A \cite{PhysRevLett.121.022504,PhysRevD.101.092001} ($\nu_\mu$ on $^{12}$C) and MicroBooNE \cite{microbooneA,microbooneB} ($\nu_\mu$ on $^{40}$Ar) collaborations.

This paper is organized as follows: In Sect.~\ref{sec:1} we present a summary of the general semi-inclusive neutrino-nucleus scattering formalism in the PWIA and provide analytic expressions of the flux-integrated fifth-differential semi-inclusive neutrino-nucleus cross sections for three different nuclear models: relativistic Fermi gas (Sect.~\ref{subsec:RFG}), independent-particle shell model (Sect.~\ref{subsec:IPSM}) and natural orbitals shell model (Sect.~\ref{subsec:NO}). Given that neutrino collaborations have presented the semi-inclusive experimental data as function of different
variables, in Sec.~\ref{sec:2} we define the three sets of observables employed in experimental analyses and give their expressions as function of the momentum and angles of the particles detected in the final state, $i.e.$, a muon and a proton in all cases considered in this work. Next, we present the comparison of our theoretical results with the available semi-inclusive data from T2K (Sect.~\ref{subsec:t2k results}), MINER$\nu$A (Sect.~\ref{subsec:minerva results}) and MicroBooNE (Sect.~\ref{subsec:microboone results}). Finally, in Sect.~\ref{sec:4} we summarize our conclusions.

\section{\label{sec:1}Semi-inclusive neutrino-nucleus reactions formalism}

Following the theoretical work started in \cite{PhysRevD.90.013014} and \cite{PhysRevC.100.044620}, we studied in \cite{PhysRevC.102.064626} the semi-inclusive neutrino-nucleus cross sections in the PWIA for three different nuclear models: the relativistic Fermi gas (RFG) model, the independent-particle shell model (IPSM) and the energy-dependent natural orbit (NO) model for $^{12}$C and $^{40}$Ar. Here we briefly summarize the main results of Ref.~\cite{PhysRevC.102.064626}.

Assuming that a neutrino of momentum $\mathbf{k}$ interacts with 
an off-shell bound nucleon of momentum $\mathbf{p_m}$ exchanging a charged boson $W$, and a lepton of momentum $\mathbf{k'}$ is detected
in coincidence with an ejected nucleon of momentum $\mathbf{p_N}$ in the final state, the semi-inclusive cross section of the process
in the factorization approximation is given by
\begin{widetext}
    \begin{align}\label{general-semics}
        \frac{d\sigma}{dk'd\Omega_{k^{'}}dp_{N}d\Omega^{L}_{N}}=\frac{(G_{F}\cos{\theta_{c}} k'p_{N})^{2}m_{N}}{8k\varepsilon'E_{N}(2\pi)^{6}}  \int_{0}^{\infty}d\mathcal{E}
        \int d^{3}p_{m} \upsilon_0\mathcal{F}^2_\chi S\bigl(p_{m},E_{m}\bigr)\nonumber\\
        \times\delta(M_{A} + k - \varepsilon' - E_{N} - \sqrt{p^{2}_{m} + M^{2}_{A-1}} - \mathcal{E}) 
        \delta (\textbf{k}- \textbf{k}' -\textbf{p}_{N} + \textbf{p}_{m}),
    \end{align}
\end{widetext}
where $G_F$ is the Fermi constant, $\theta_c$ is the Cabibbo angle, and $(\varepsilon',\Omega_{k^{'}})$ and  $(E_N,\Omega^{L}_{N})$ are the energies and scattering angles of the final lepton and nucleon,
respectively. $S\bigl(p_{m},E_{m}\bigr)$ is the nuclear spectral function that describes the probability of finding a nucleon in a nucleus $A$ with
given momentum (called the missing momentum $p_m$) and with a given excitation energy of the residual nuclear system $A-1$ (called the missing energy $E_m$).
The kinematic factor $\upsilon_0$ and the reduced single nucleon cross section $\mathcal{F}^2_\chi$ are defined in Appendix A of \cite{PhysRevC.102.064626} and
$\mathcal{E}$ is the difference in recoil energies 
between the residual nucleus with invariant mass $W_{A-1}$ and one with minimum invariant
mass $M_{A-1}$. Note that \eq{general-semics} depends on the neutrino momentum $k$, the final lepton variables ($k', \theta_l$, $\phi_l$) and also the final nucleon variables
($p_N$, $\theta_N^L$, $\phi_N^L$). For convenience the neutrino direction is chosen as the $z$-axis, therefore $\theta_N^L$ is the angle between the final nucleon and the
incoming neutrino and $\phi_N^L$ is the angle between the scattering plane (defined by $\mathbf{k}$ and $\mathbf{k'}$) and the reaction plane (defined by $\mathbf{k}$ and $\mathbf{p_N}$).
The two $\delta$-functions that appear in \eq{general-semics} guarantee the conservation of energy and momentum in the process.

In what follows we briefly summarize the three different nuclear models considered in this work, namely the RFG, IPSM and NO,
and obtain the analytical expressions for the semi-inclusive neutrino-nucleus cross section in each model, providing the explicit form of the spectral function $S\bigl(p_{m},E_{m}\bigr)$. The normalization of the spectral function is chosen to be
\begin{equation}\label{norm-sf}
    \mathcal{N}=\frac{1}{(2\pi)^{3}}\int_{0}^{\infty}dp_{m}p^{2}_{m}\int_{0}^{\infty}dE_{m}S(p_{m},E_{m})
\end{equation} 
with $\mathcal{N}$ the number of nucleons that are active in the scattering, $i.e.$, the number of neutrons for the case of neutrino scattering (CC$_{\nu}$)
and the number of protons for antineutrinos (CC$_{\bar{\nu}}$). 

\subsection{\label{subsec:RFG} Relativistic Fermi Gas (RFG)}
This model consists in describing the nucleus as an infinite gas of free relativistic nucleons that, in the nuclear ground state, occupy all the levels up to the Fermi momentum $k_F$
while the levels above that are empty. The Fermi momentum is the only free parameter of the model. It is usually fitted to the width of the quasielastic peak in electron scattering data
and varies with the nucleus. As proposed in \cite{PhysRevC.100.044620}, we consider a shift in the RFG energies by a constant quantity in such a way that the last occupied level
in the Fermi sea coincides with the separation energy -$E_s$, defined as the minimum energy necessary to remove a nucleon from the nucleus. This introduces a second parameter in the model, $E_s$, fitted to the position of the quasielastic peak. By doing this, the nucleons in the RFG model are no longer 
on-shell because their free energy is changed as 
\begin{equation}
    E=\sqrt{p_m^2+m_N^2} \ \ \longrightarrow \ \ E-\left(E_F + E_s\right)
\end{equation}
with $E_F = \sqrt{k_F^2 + m_N^2}$ the Fermi energy. For this model, the normalized spectral function is \cite{PhysRevC.100.044620}
\begin{equation}
    S_{RFG}(p_m, \mathcal{E})=\frac{3(2\pi)^{3}\mathcal{N}}{k^{3}_{F}}\Theta(k_{F} - p_{m})\delta\bigl(\mathcal{E} - E_F + E\bigr), 
\end{equation}
where the Pauli principle is explicitly imposed by the $\Theta$-function and the energy conservation by the $\delta$-function.

Given a
normalized neutrino flux $P(k)$, the flux-averaged semi-inclusive neutrino-nucleus cross section for the RFG model is
\begin{widetext}
    \begin{align}\label{semi-inclusive_RFG}
    \left <\frac{d\sigma}{dk'd\Omega_{k'}dp_{N}d\Omega^{L}_{N}}\right >&=\frac{3\mathcal{N}(G_{F}\cos{\theta_{c}}m_{N}k'p_{N})^{2}}{8(2\pi k_{F})^{3}\varepsilon'E_{N}}  \frac{P(k_{0})}{k_{0}}\frac{\upsilon_{0}\mathcal{F}^{2}_{\chi}}{E_{B} - p_{B}\cos{\theta_{B}}}\Theta(k_{F} - {p}_{m}) \Theta(p_N-k_F) \, ,
    \end{align}
\end{widetext}
where we have defined the following auxiliary variables:
\begin{align}
    \textbf{p}_{B} =\, &\textbf{k}' + \textbf{p}_{N}, \\
    E_{B} =\, &\varepsilon' + E_{s} + E_{F} + E_{N} - m_N, \\
    p_B\cos{\theta_{B}} =\, &k'\cos{\theta_{l}} + p_{N}\cos{\theta^{L}_{N}}.
\end{align}
Then, in the RFG model, the neutrino momentum $k_0$ is fixed by energy conservation to
\begin{equation}
    k_{0} = \frac{E^{2}_{B} - p^{2}_{B}-m^{2}_{N}}{2\bigl(E_{B} - p_{B}\cos\theta_{B} \bigr)}
\end{equation}
and the missing momentum $p_m$ by momentum conservation to
\begin{align}\label{pmiss_RFG}
    &&{p}^{2}_{m}= k^{2}_0 - 2k'k_0\cos{\theta_{l}} + k'^{2} + p^{2}_{N} -2k_{0}p_{N}\cos{\theta^{L}_{N}}  \nonumber\\
    &&+ 2k'p_{N}(\cos{\theta_{l}}\cos{\theta^{L}_{N}} 
    + \sin{\theta_{l}}\sin{\theta^{L}_{N}}\cos{\phi^{L}_{N}}).
\end{align}
Note that both $k_0$ and $p_m$ are completely determined by the final state kinematics.

\subsection{\label{subsec:IPSM}Independent-particle shell model (IPSM)}

In the IPSM the bound nucleons are described by wave functions that are solutions of the Dirac-Hartree equation with real scalar and vector potentials and occupy discrete energy levels $-E_{nlj}$. The spectral function of this model is \cite{PhysRevC.100.044620}
\begin{equation}\label{ipsm-spectralfunction}
    S_{\textrm{IPSM}}(p_{m},\mathcal{E})=\sum_{n,l,j}(2j +1)n_{nlj}(p_{m})\delta(\mathcal{E} + E_{s} - E_{nlj})\, ,
\end{equation} 
where $n_{nlj}(p_m)$ is the momentum distribution of a single nucleon in a shell characterized by the quantum numbers $nlj$. The flux-averaged semi-inclusive neutrino-nucleus 
cross section for this model is then 
\begin{align}\label{semi-inclusive_IPSM}
    &\left <\frac{d\sigma}{dk'd\Omega_{k'}dp_{N}d\Omega^{L}_{N}}\right > = \frac{(G_{F}\cos{\theta_{c}}k'p_{N})^{2}m_{N}}{8(2\pi)^{6}\varepsilon'E_{N}}  \sum_{n,l,j}(2j +1)\nonumber\\
    &\qquad\qquad\qquad\quad\times\frac{P(k_{0nlj})}{k_{0nlj}}\upsilon_{0}\mathcal{F}^{2}_{\chi}n_{nlj}(p_{m})
\end{align}
with the neutrino momentum
\begin{equation}
    k_{0nlj} = \varepsilon' + E_{N} - m_{N} + E_{nlj} 
\end{equation}
and the missing momentum
\begin{align}
    p^{2}_{m}&= k^{2}_{0nlj} + k'^{2} + p^{2}_{N} -2k_{0nlj}k'\cos{\theta_{l}} -2k_{0nlj}p_{N}\cos{\theta^{L}_{N}} \nonumber \\ &+ 2k'p_{N}(\cos{\theta_{l}}\cos{\theta^{L}_{N}}+ \sin{\theta_{l}}\sin{\theta^{L}_{N}}\cos{\phi^{L}_{N}})
\end{align}
fixed by the energy and momentum conservation. Notice that, unlike the RFG case, for the IPSM the final state kinematics does not correspond to a definite neutrino energy. This is a consequence of the shell structure of the spectral function.

\subsection{\label{subsec:NO} Natural orbitals shell model (NO)}

This model is similar to the IPSM but it takes into account nucleon-nucleon (NN) correlations and the smearing of the energy eigenstates. It employs natural orbitals, $\psi_{i}(r)$, defined as the complete orthonormal set of single-particle wave functions that diagonalize the one-body density matrix (OBDM)~\cite{PhysRev.97.1474}:
\begin{equation}\label{OBDM}
    \rho(\textbf{r},\textbf{r}' )=\sum_{i} N_{i}\psi_{i}^{*}(\textbf{r})\psi_{i}(\textbf{r}')\, ,
\end{equation}
where the eigenvalues $N_{i} (0\le N_{i} \le 1, \sum_{i}N_{i}=A)$ are the natural occupation numbers and $A$ is the mass number.

The NO single-particle wave functions are used to obtain the occupation numbers and the wave functions in momentum space, {\it i.e.,} 
the momentum distributions, and from them the spectral function that is given by~\cite{PhysRevC.89.014607} 
\begin{equation}\label{spectral-sofia}
    S_{\textrm{NO}}(p_{m},\mathcal{E})=\frac{1}{2\pi A}\sum_{i}(2j_{i} + 1)N_{i}|\psi_{i}(p_{m})|^{2}L_{\Gamma_{i}}(\mathcal{E}-\mathcal{E}_{i})\, ,
\end{equation}
where the dependence upon the energy is given by the Lorentzian function:
\begin{equation}\label{lorentzian}
    L_{\Gamma_{i}}(\mathcal{E}-\mathcal{E}_{i})=\frac{1}{2\pi}\frac{\Gamma_{i}}{(\mathcal{E} - \mathcal{E}_{i})^{2} + (\Gamma_{i}/2)^{2}}
\end{equation}
with $\Gamma_{i}$ the width for a given single-particle state and $\mathcal{E}_{i}$ the energy eigenvalue of the state. 

The flux-averaged semi-inclusive neutrino-nucleus cross section in this model is given by 
\begin{align}\label{seminclusive-sofia}
    &\left <\frac{d\sigma}{dk'd\Omega_{k'}dp_{N}d\Omega^{L}_{N}}\right >=\int d\mathcal{E} \frac{(G_{F}\cos{\theta_{c}}k' p_{N})^{2}m_{N}P(k)}{8k\varepsilon'E_{N}(2\pi)^{7}A} \nonumber \\
    &\qquad\quad \times\sum_{i}(2j_{i} +1)N_{i}|\psi_{i}(p_{m})|^{2}L_{\Gamma_{i}}(\mathcal{E} - \mathcal{E}_{i})\upsilon_{0}\mathcal{F}^{2}_{\chi}\, ,
\end{align}
where the neutrino momentum, for a given excitation energy $\mathcal{E}$, is
\begin{equation}\label{neutrino-momentum-sofia}
    k=E_{s} + E_{N} + \varepsilon' - m_{N} + \mathcal{E}
\end{equation}
and
\begin{align}\label{missing-momentum-sofia}
p^{2}_{m}&= k^{2} + k'^{2} + p^{2}_{N} -2kk'\cos{\theta_{l}} -2kp_{N}\cos{\theta^{L}_{N}} \nonumber \\ &+ 2k'p_{N}(\cos{\theta_{l}}\cos{\theta^{L}_{N}}+ \sin{\theta_{l}}\sin{\theta^{L}_{N}}\cos{\phi^{L}_{N}})\, .
\end{align}
Note that in this case the integral over $\mathcal{E}$ in \eq{seminclusive-sofia} has to be performed numerically because, unlike the IPSM, the single-particle energies are not discrete. 

\begin{figure}
	\centering
	\includegraphics[width=0.4\textheight]{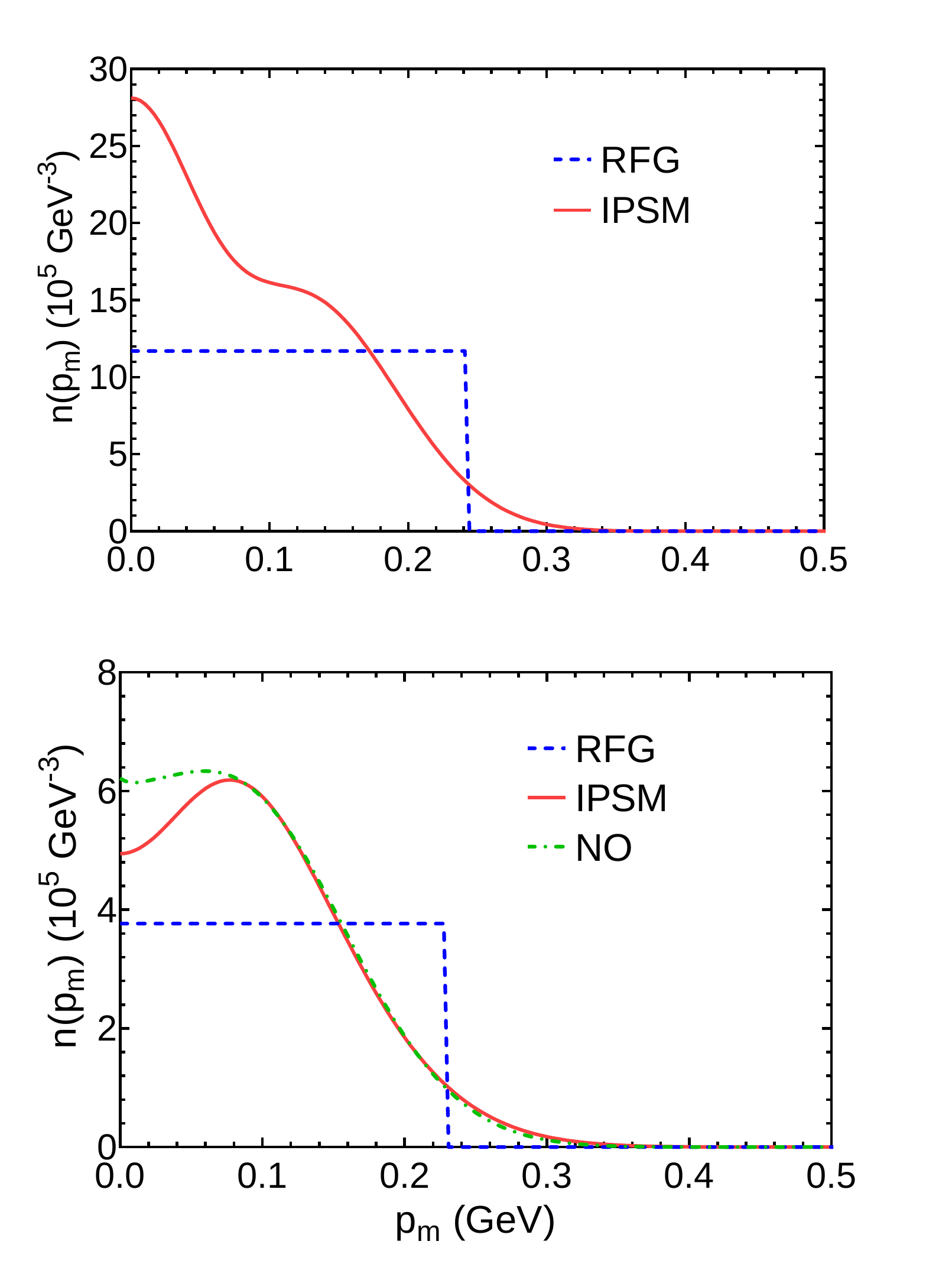}
	\caption{\label{fig:momentum distributions}Momentum distributions for $^{40}$Ar (top) and $^{12}$C (bottom) normalized according to \eq{norm-sf} for the three different nuclear models considered in this work, namely RFG (blue dashed), IPSM (red solid) and NO (green dot-dashed).}
\end{figure}

Momentum distributions for the three nuclear models considered are shown in Fig.~\ref{fig:momentum distributions} (bottom) for $^{12}$C. In the case of $^{40}$Ar, only two nuclear models are considered, namely RFG and IPSM, and their respective momentum distributions are also shown in Fig.~\ref{fig:momentum distributions} (top). Notice the difference between the IPSM and NO predictions for $^{12}$C in the region of low missing momentum.

\section{\label{sec:2}Experimental observables}

The main objective of this work is to compare all the available semi-inclusive experimental data for different experiments, namely, T2K, MINER$\nu$A and MicroBooNE, with theoretical predictions in the PWIA using different nuclear models.
The kinematics of the outgoing muon and proton for semi-inclusive CC0$\pi$ events is completely characterized by the independent variables $(k', \theta_l, p_N, \theta_N^L, \phi_N^L)$, which will be called natural variables (NV).
In addition to these variables, we 
also introduce two more sets of variables used in the experimental analyses: the transverse kinematic imbalances (TKI) \cite{PhysRevC.94.015503,doi:10.7566/JPSCP.12.010032,PhysRevD.101.092001} and the inferred variables (IV) \cite{PhysRevD.98.032003}, which will be defined in the following sections.

\subsection{\label{subsec:natural variables}Natural variables (NV)}

\begin{figure}[!htbp]
    \centering
    \includegraphics[width=0.45\textwidth]{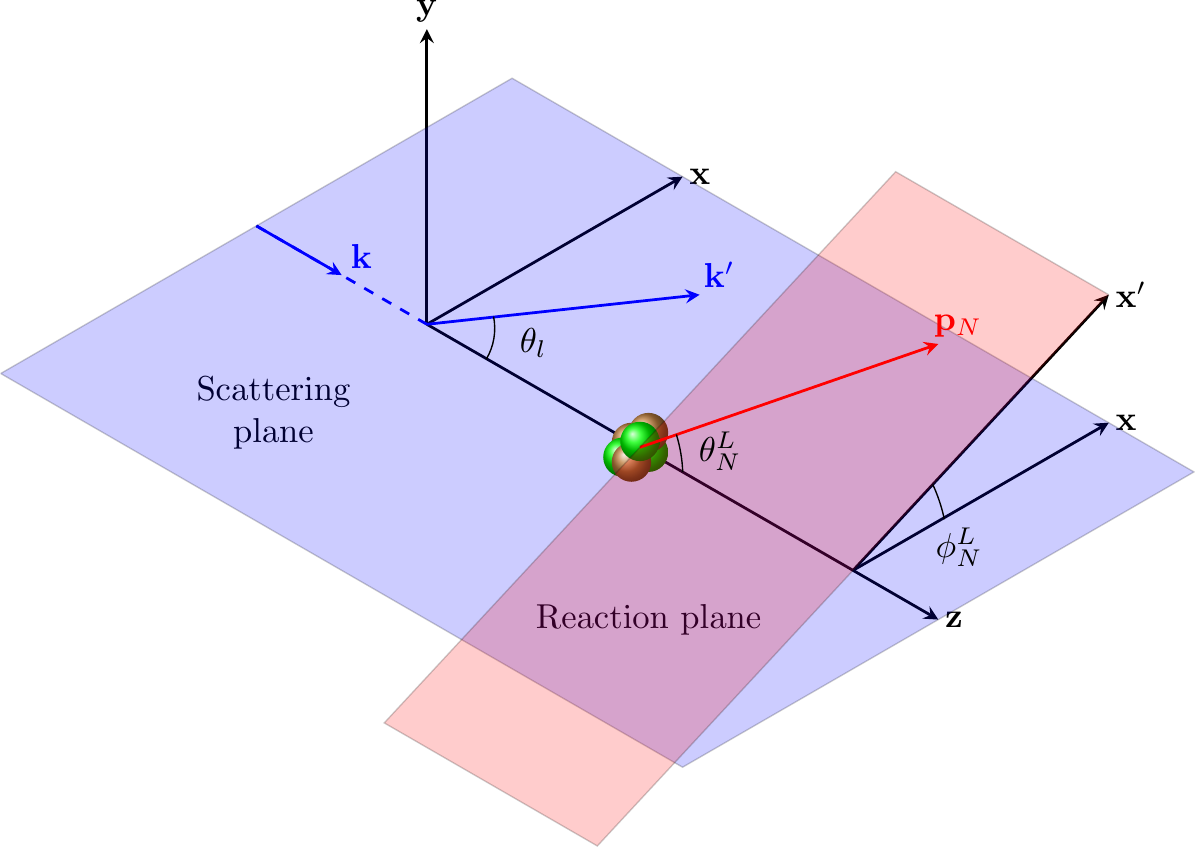}
    \caption{\label{fig:nv}Schematic representation of the definition of the natural variables: $k'$, $\theta_l$, $p_N$, $\theta_N^L$ and $\phi_N^L$. The incoming neutrino ($\mathbf{k}$) and the final lepton ($\mathbf{k}'$) are contained in the 
    scattering plane, while the reaction plane contains the incoming neutrino and the ejected nucleon. }
\end{figure}

The first and more straightforward way to characterize the final state is using the so-called natural variables (NV) defined in Fig.~\ref{fig:nv}. Taking the $z$-axis as the incoming neutrino direction, the final lepton momentum ($\bf k'$)
forms an angle $\theta_l$ with the initial neutrino direction ($\bf k$) and 
the two vectors define the scattering plane ($xz$-plane). After the interaction with the nucleus, a nucleon is ejected with momentum 
$\bf p_N$ forming an angle $\theta_N^L$ with
the initial neutrino direction. The final nucleon is contained in a plane called reaction plane ($x'z$-plane) which is rotated by an angle $\phi_N^L$ with respect to the scattering plane. 

This set of variables will be used in the results section to analyze both semi-inclusive and inclusive CC0$\pi$; the latter are obtained by integrating \eq{general-semics} over the final nucleon variables.

The three-momenta defined in the frame shown in Fig.~\ref{fig:nv} are
\begin{align}
    \mathbf{k} &= k\mathbf{e}_z, \nonumber \\
    \mathbf{k}' &= k'\bigl(\sin{\theta_l}\mathbf{e}_x + \cos{\theta_l}\mathbf{e}_z\bigr), \nonumber \\
    \mathbf{p}_N &= p_N\bigl(\sin{\theta_N^L}\cos{\phi_N^L}\mathbf{e}_x + \sin{\theta_N^L}\sin{\phi_N^L}\mathbf{e}_y + \cos{\theta_N^L}\mathbf{e}_z\bigr).
\end{align}
 
\subsection{\label{subsec:stv variables}Transverse kinematic imbalances (TKI)}

\begin{figure}[!htbp]
    \centering
    \includegraphics[width=0.48\textwidth]{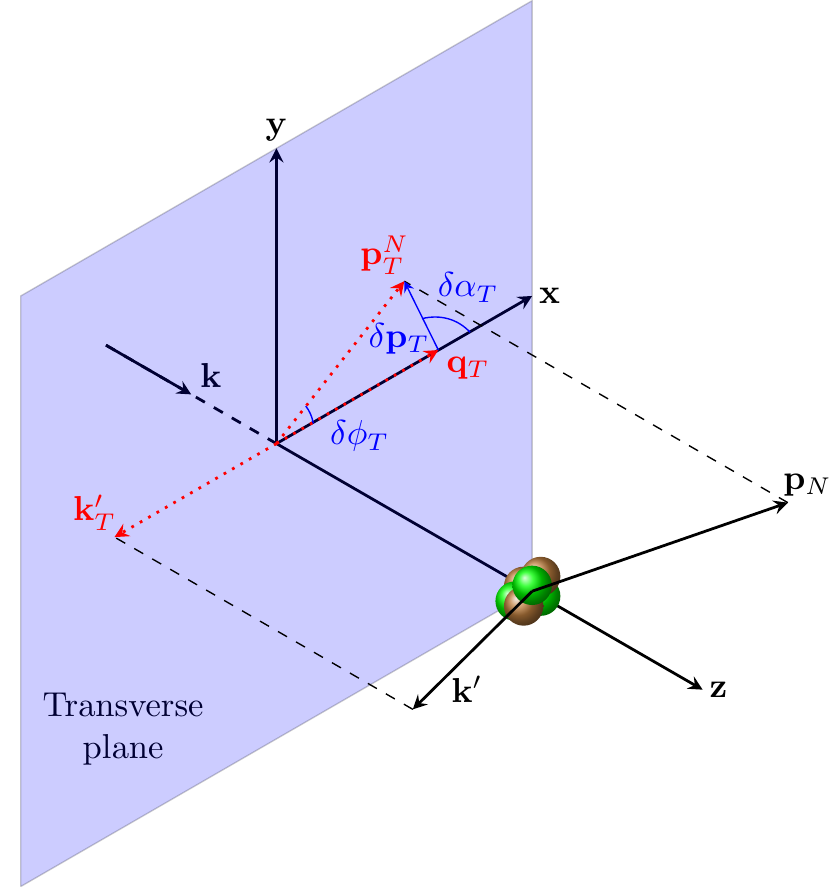}
    \caption{\label{fig:stv}Scheme showing transverse kinematic imbalances (TKI): $\delta p_T$, $\delta\alpha_T$ and $\delta\phi_T$. The final lepton and nucleon momenta are projected on the plane perpendicular to the neutrino direction ($xy$-plane or transverse plane).
    The transverse component of the transferred momentum ($\mathbf{q}_T$) equals $-\mathbf{k}_T'$ and defines the $x$-axis.}
\end{figure}

The transverse kinematic imbalances (TKI) \cite{PhysRevC.94.015503} are designed to enhance some nuclear effects, and therefore discriminate between different models, with minimal dependence on the neutrino energy. In particular, the use of TKI can help in disentangling effects linked to final state interactions (FSI), initial state correlations and/or multinucleon excitations (2p2h).
%
They are defined by projecting the final lepton and the ejected nucleon momenta on the plane perpendicular to the neutrino direction (transverse plane) as can be seen in Fig.~\ref{fig:stv}.

More specifically, the vector magnitude of the momentum imbalance ($\delta p_T$) and the two angles ($\delta\alpha_T$ and $\delta\phi_T$) are:
\begin{align}\label{stv definition}
    \delta p_T = \bigl|\mathbf{k}_T' + \mathbf{p}^N_T \bigr|, \nonumber \\
    \delta\alpha_T = \text{arc\,cos}\frac{-\mathbf{k}_T' \cdot \mathbf{\delta p}_T}{k_T'\delta p_T}, \nonumber\\
    \delta\phi_T = \text{arc\,cos}\frac{-\mathbf{k}_T' \cdot \mathbf{p}_T^N}{k_T'p_T^N},
    \end{align}

where $\mathbf{k}_T'$ and $\mathbf{p}_T^N$ are, respectively, the projections of the final lepton and nucleon momentum on the transverse plane (if the neutrino direction is taken as the $z$-axis, then the projections only have components in the $xy$-plane). In the PWIA, for which $\mathbf{k}'+\mathbf{p}_N=\mathbf{k}+\mathbf{p}_m$, $\delta \mathbf{p}_T$ is the transverse component of the initial nucleon momentum and $\delta\alpha_T$ the angle between the transverse projections of the initial nucleon momentum and the transferred momentum $\mathbf{q}$. Also, in absence of FSI, the $\delta\alpha_T$ distribution is expected to be flat.

The T2K and MINER$\nu$A collaborations measured single differential cross sections with respect to TKI defined in \eq{stv definition}. These can be calculated by integrating the sixth-differential semi-inclusive cross section in \eq{general-semics}, given in terms of the NV, over five of the six variables, after performing the appropriate change of variables. Hence, in order to compare with these data, we need to connect the two sets of variables and evaluate the necessary Jacobians. The details of these transformations are given in the Appendix \ref{sec:app}, leading to the following expressions:

\begin{widetext}    
    \begin{eqnarray}
    \label{sdcs deltapt}
        \frac{d\sigma}{d\delta p_T} &=& 2\pi\int_{\theta_l^{min}}^{\theta_l^{max}}d\theta_l\sin{\theta_l}\int_{k'_{min}}^{k'_{max}}dk'\int_{\theta_N^{Lmin}}^{\theta_N^{Lmax}}d\theta_N^L\sin{\theta_N^L}\int_{\phi_N^{Lmin}}^{\phi_N^{Lmax}}d\phi_N^L
       \left < d^5\sigma \right >\mathcal{J}_p\Theta(p_N - p_N^{min})\Theta(p_N^{max}-p_N),
\\
\label{sdcs deltaalphat}
        \frac{d\sigma}{d\delta \alpha_T} &=& 2\pi\int_{\theta_l^{min}}^{\theta_l^{max}}d\theta_l\sin{\theta_l}\int_{k'_{min}}^{k'_{max}}dk'\int_{p_N^{min}}^{p_N^{max}}dp_N\int_{\phi_N^{Lmin}}^{\phi_N^{Lmax}}d\phi_N^L
     \left < d^5\sigma \right >\mathcal{J}_\alpha\sin{\theta_N^L}\Theta(\theta_N^L - \theta_N^{Lmin})\Theta(\theta_N^{Lmax}-\theta_N^L),\\
\label{sdcs deltaphit}
        \frac{d\sigma}{d\delta \phi_T} &=& 2\pi\int_{\theta_l^{min}}^{\theta_l^{max}}d\theta_l\sin{\theta_l}\int_{k'_{min}}^{k'_{max}}dk'\int_{p_N^{min}}^{p_N^{max}}dp_N\int_{\theta_N^{Lmin}}^{\theta_N^{Lmax}}d\theta_N^L\left < d^5\sigma \right >\sin{\theta_N^L},
	\end{eqnarray}
\end{widetext}
where $\left< d^5\sigma \right >$ is the flux-averaged fifth-differential semi-inclusive cross section

\begin{equation}
\left< d^5\sigma \right > = \left< \frac{d\sigma}{dk'd\Omega_{k^{'}}dp_{N}d\Omega^{L}_{N}} \right>
\end{equation}
and the Jacobians $ \mathcal{J}_p $ and $ \mathcal{J}_\alpha$ have been defined, respectively, in \eq{jacopt} and \eq{jacoalpha}.
The integration limits are either dictated by the kinematics or imposed by the experimental cuts, as will be shown in results section.

\subsection{\label{subsec:inferred variables}Inferred variables (IV)}

The T2K collaboration also measured \cite{PhysRevD.98.032003} single differential cross sections as function of the so-called inferred variables, which compare the momentum and angle of the ejected proton with the proton
kinematics inferred from the measured final muon kinematics under the so-called QE~hypothesis, $i.e.$ initial nucleon at rest. In this approximation, the neutrino energy and the final proton momenta are defined as
\begin{equation}\label{neutrino inferred energy}
    E_\nu = \frac{m_p^2 - m_{l}^2 + 2E_l(m_n - E_b) - (m_n - E_b)^2}{2[m_n - E_b - E_l + k'\cos{\theta_l}]}
\end{equation}
and
\begin{equation}\label{pn_inferred}
    \mathbf{p}_N^{\text{inf}} = \bigl(-k'\sin{\theta_l}, 0 , -k'\cos{\theta_l} + E_\nu   \bigr),
\end{equation}
where the $z$-axis corresponds to the neutrino direction; $m_n$, $m_p$ and $m_l$ are the neutron, proton and muon masses and $E_b$ and $E_l$ are the nuclear binding energy fixed to 25~MeV for $^{12}$C and the muon energy.
Then, when a muon and (at least) one proton are measured in the final state we can define three observables:
\begin{align}\label{inferred definition}
    \Delta p &= \left|\mathbf{p}_N\right| - \left|\mathbf{p}_N^{\text{inf}}\right|,\nonumber\\
    \Delta \theta &= \theta_N^L - \theta_N^{\text{inf}},\nonumber\\
    \left|\Delta \mathbf{p}\right| &= \left| \mathbf{p}_N - \mathbf{p}_N^{\text{inf}} \right|
\end{align}
with $\mathbf{p}_N$ the three-momentum of the ejected nucleon, $\theta_N^L$ the angle between the final nucleon and the direction of the incoming neutrino, and $\theta_N^{\text{inf}}$ the angle between the neutrino direction and the three-momentum
of the ejected nucleon in the QE hypothesis defined as
\begin{equation}\label{thetan_inferred}
    \cos{\theta_N^{\text{inf}}} = \frac{\mathbf{p}_N^{\text{inf}}\cdot\mathbf{z}}{\left| \mathbf{p}_N^{\text{inf}}\right|} = \frac{-k'\cos{\theta_l} + E_\nu}{\left| \mathbf{p}_N^{\text{inf}}\right|}.
\end{equation}

The definition \eqref{inferred definition} of $\left|\Delta \mathbf{p}\right|$ can be expressed as a second degree equation for $p_N$ in the form $p_N^2 +2b'p_N + c' = 0$ as follows
\begin{widetext}    
\begin{align}\label{modvecp second degree}
    p_N^2 + 2p_N\bigl( k'\sin{\theta_N^L}\cos{\phi_N^L}\sin{\theta_l} -E_\nu\cos{\theta_N^L} + k'\cos{\theta_N^L}\cos{\theta_l}\bigr) + 
    {k'}^2\sin^2{\theta_l} + \bigl(k'\cos{\theta_l} - E_\nu\bigr)^2 - \left|\Delta\mathbf{p}\right|^2 = 0.
\end{align}
\end{widetext}
Notice that, according to \eq{pn_inferred} and \eq{thetan_inferred}, the definition of the inferred proton kinematics relies on the same QE expression used in the estimation of neutrino energy in oscillations measurements.
Consequently, the observed deviations from the expected proton inferred kinematic imbalance could provide hints of the biases that may be caused from the mismodeling of nuclear effects in neutrino oscillations measurements at T2K.

For the inferred variables, the single differential cross sections can be defined following the same procedure used with TKI in the previous section, yielding
\begin{widetext}    
    \begin{align}\label{sdcs Deltap}
        \frac{d\sigma}{d\Delta p} &= 2\pi\int_{\theta_l^{min}}^{\theta_l^{max}}d\theta_l\sin{\theta_l}\int_{k'_{min}}^{k'_{max}}dk'\int_{\theta_N^{Lmin}}^{\theta_N^{Lmax}}d\theta_N^L\sin{\theta_N^L}\int_{\phi_N^{Lmin}}^{\phi_N^{Lmax}}d\phi_N^L
        \left < d^5\sigma \right > \Theta(p_N - p_N^{min})\Theta(p_N^{max}-p_N), \nonumber\\[2.5mm] 
        \frac{d\sigma}{d\Delta\theta} &= 2\pi\int_{\theta_l^{min}}^{\theta_l^{max}}d\theta_l\sin{\theta_l}\int_{k'_{min}}^{k'_{max}}dk'\int_{p_N^{min}}^{p_N^{max}}dp_N\int_{\phi_N^{Lmin}}^{\phi_N^{Lmax}}d\phi_N^L
        \left < d^5\sigma \right >\sin{\theta_N^L}\Theta(\theta_N^L - \theta_N^{Lmin})\,\Theta(\theta_N^{Lmax}-\theta_N^L), \nonumber\\[2.5mm] 
        \frac{d\sigma}{d\left|\Delta \mathbf{p}\right|} &= 2\pi\int_{\theta_l^{min}}^{\theta_l^{max}}d\theta_l\sin{\theta_l}\int_{k'_{min}}^{k'_{max}}dk'\int_{\theta_N^{Lmin}}^{\theta_N^{Lmax}}d\theta_N^L\sin{\theta_N^L}\int_{\phi_N^{Lmin}}^{\phi_N^{Lmax}}d\phi_N^L
        \left < d^5\sigma \right > \mathcal{J}_{\Delta \mathbf{p}}\Theta(p_N - p_N^{min})\Theta(p_N^{max}-p_N),
    \end{align}
\end{widetext}
with $\mathcal{J}_{\Delta \mathbf{p}}$ the jacobian of the variable change $p_N \rightarrow \left|\Delta \mathbf{p}\right|$
\begin{equation}
    \mathcal{J}_{\Delta \mathbf{p}} = \biggl| \frac{\partial p_N}{\partial\left|\Delta \mathbf{p}\right|} \biggr|_{(\theta_N,\phi_N)} = \biggl|\frac{p_N + b'}{\Delta \mathbf{p}}\biggr|.
\end{equation}
As it happened for TKI, also in this case the second degree equation \eqref{modvecp second degree} can have (see Appendix~\ref{sec:app}) zero, one or two valid solutions in the variable $p_N$  (namely belonging to the interval $[p_N^{min},p_N^{max}]$). In the latter case the contributions from both solutions must be summed.

\section{\label{sec:3}Results}

In this section we compare our theoretical results based on the PWIA with CC0$\pi$ measurements of three neutrino collaborations: T2K, MINER$\nu$A and MicroBooNE.
For each experiment, we compare our predictions with inclusive single or double differential cross sections as function of the muon momentum and scattering angle. Next, we show the semi-inclusive cross section results (CC0$\pi$1p where a muon and a proton are detected in the final state) as function of NV, IV or TKI depending on the available data from each experiment.

\subsection{\label{subsec:t2k results}T2K}

The T2K data-sample \cite{PhysRevD.98.032003} includes CC0$\pi$ inclusive cross sections without protons in the final state as function of final muon variables and CC0$\pi$1p semi-inclusive cross sections with at least one proton and one muon in the final state as function of NV, IV and TKI. The phase-space restrictions applied to the analyses as function of each set of variables are summarized in Table~\ref{table:T2K constrains}.

\begin{table*}[!t]
    \centering
    \begin{tabular}{cccccccccccccccc}
    \hline
    \toprule\toprule
        \Mark{T2K}              &  &   $k'$  &   & & $\cos{\theta_l}$ & &    &  $p_N$  &  && $\cos{\theta_N^L}$& && $\phi_N^L$ &\\\midrule

    Inclusive NV      &  &    -    &   &  &        -        &  & & $< 0.5$ GeV & &  &       -         & && -&\\ \midrule
    Semi-inclusive NV &  &    -    &  &    &      -         & &  &$> 0.5$ GeV&  &&         -         &&&-&\\ \midrule
    TKI               & & $> 0.25$ GeV & &  &      $> -0.6$   &  &  &0.45-1.0 GeV& &  &    $> 0.4$    &&&-&  \\ \midrule
    IV                &  &    -   &    &   &       -         & &  &$> 0.45$ GeV& & &     $> 0.4$      &&&-&\\ 
    \bottomrule\bottomrule
    \hline
\end{tabular}
\caption{\label{table:T2K constrains}Final muon and proton phase-space restrictions applied to the CC0$\pi$ data shown by T2K collaboration in \cite{PhysRevD.98.032003}.}
\end{table*}

\begin{figure*}[!htbp]
		\centering
		\makebox[\textwidth]{\includegraphics[width=\textwidth,height=0.85\paperheight]{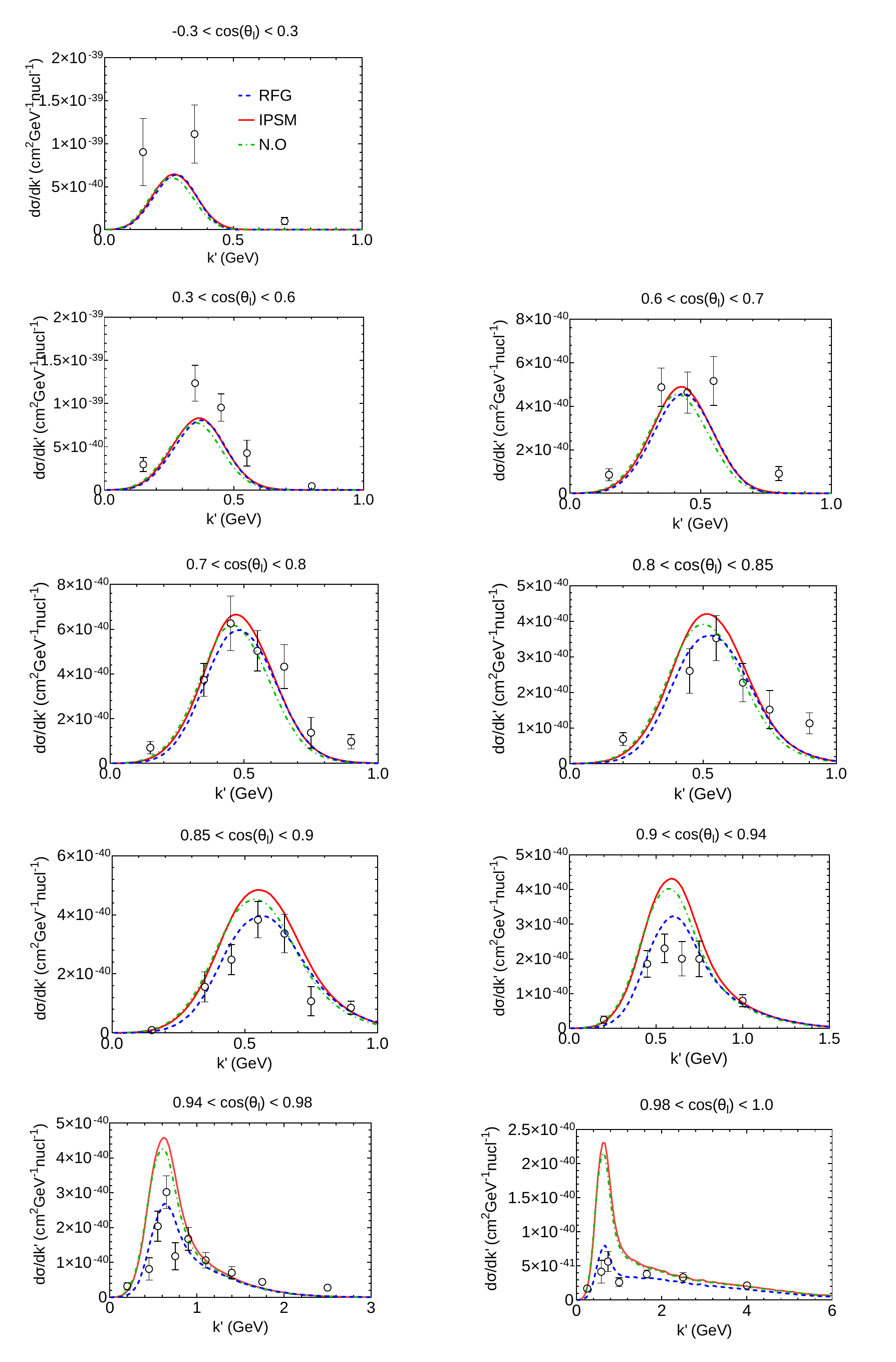}}
		\caption{\label{fig:T2K inclusive}The T2K CC0$\pi$ inclusive $\nu_\mu-^{12}$C cross sections without protons in the final state with momenta above 0.5 GeV as function of final muon kinematics for different nuclear models. Data taken from \cite{PhysRevD.98.032003}.}
\end{figure*}

Starting with the CC0$\pi$ data, in Fig.~\ref{fig:T2K inclusive} we show the flux-averaged differential inclusive cross sections for $^{12}$C evaluated for the three nuclear models considered, RFG (blue dashed), IPSM (solid red) and NO (green dot-dashed).  Note that, as specified in Table \ref{table:T2K constrains}, only the outgoing protons with momentum $p_N < 0.5$ GeV contribute to the experimental signal. Accordingly, only these events are included in the theoretical calculations. Despite the very different momentum distributions for the three models, particularly for RFG, the corresponding inclusive cross sections are rather similar for $\cos\theta_l$ smaller than about 0.8, becoming more and more different from each other as the scattering angle approaches zero ($i.e.$, small  transferred energy), where the IPSM and NO results are much higher than the data whereas the RFG ones stay close to the data. As discussed in \cite{PhysRevC.101.015503, Megias_2018}, the PWIA approach fails to describe lepton-nucleus scattering reactions at low values of the momentum and energy transfers. This is a consequence of the lack of orthogonality between the bound and free nucleon wave functions, and of the large effects associated to the overlap between the initial and final states. As already noticed, for the RFG model the cross section at very small scattering angles is reduced and compares better with the data. This is the effect of Pauli blocking, which is by definition included in the RFG model and implies that the ejected nucleon must obey $p_N > k_F$. We expect that the orthogonalization of the initial and final nuclear wave functions, as well as the implementation of FSI, will bring the IPSM and NO results closer to the data even for the smaller scattering angles. Work along these lines is in progress.

\begin{figure}[!htbp]
	\captionsetup[subfigure]{labelformat=empty} 
	\centering
	\subfloat[]{\includegraphics[width=0.52\textwidth]{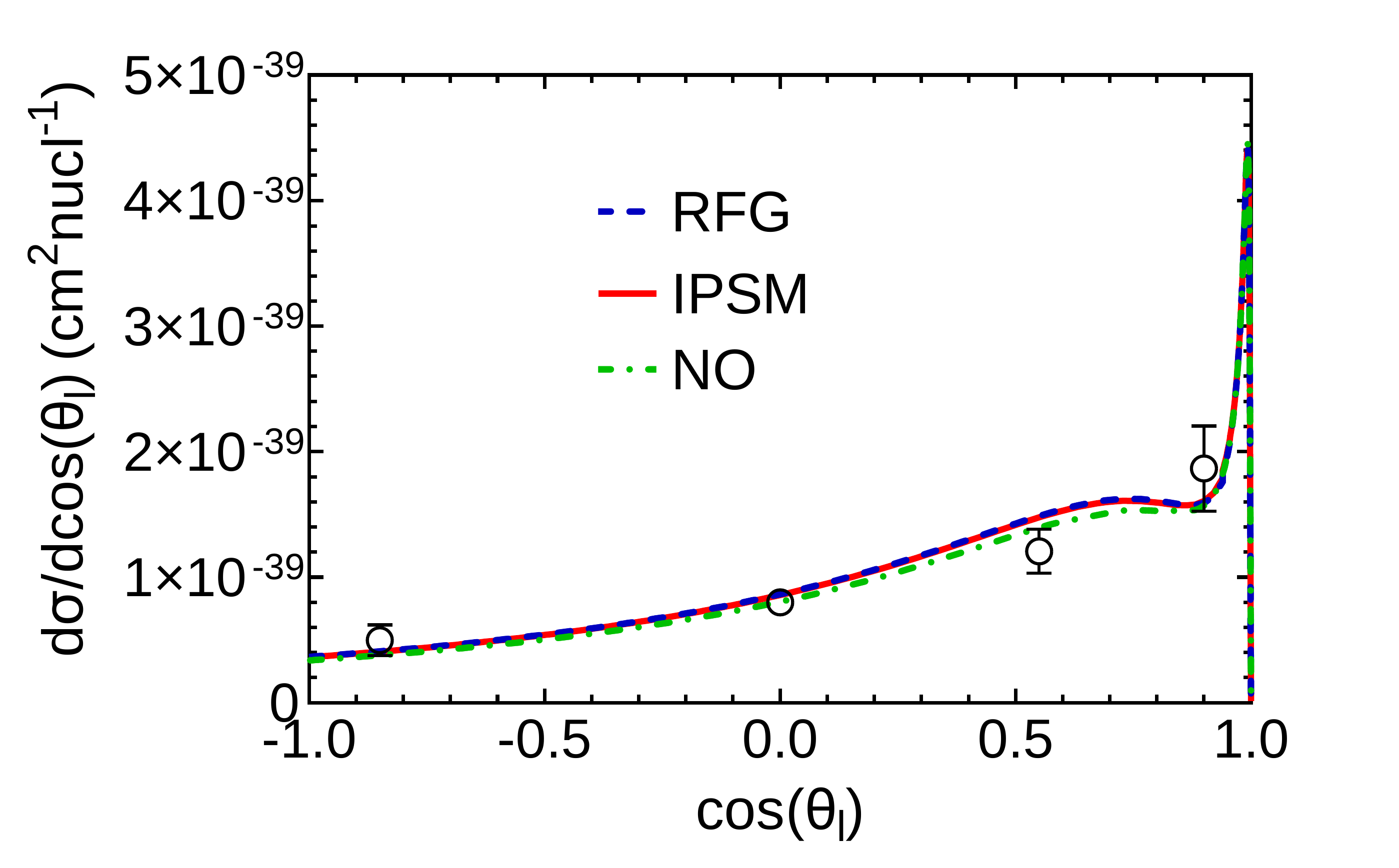}}
	\quad
	\subfloat[]{\includegraphics[width=0.52\textwidth]{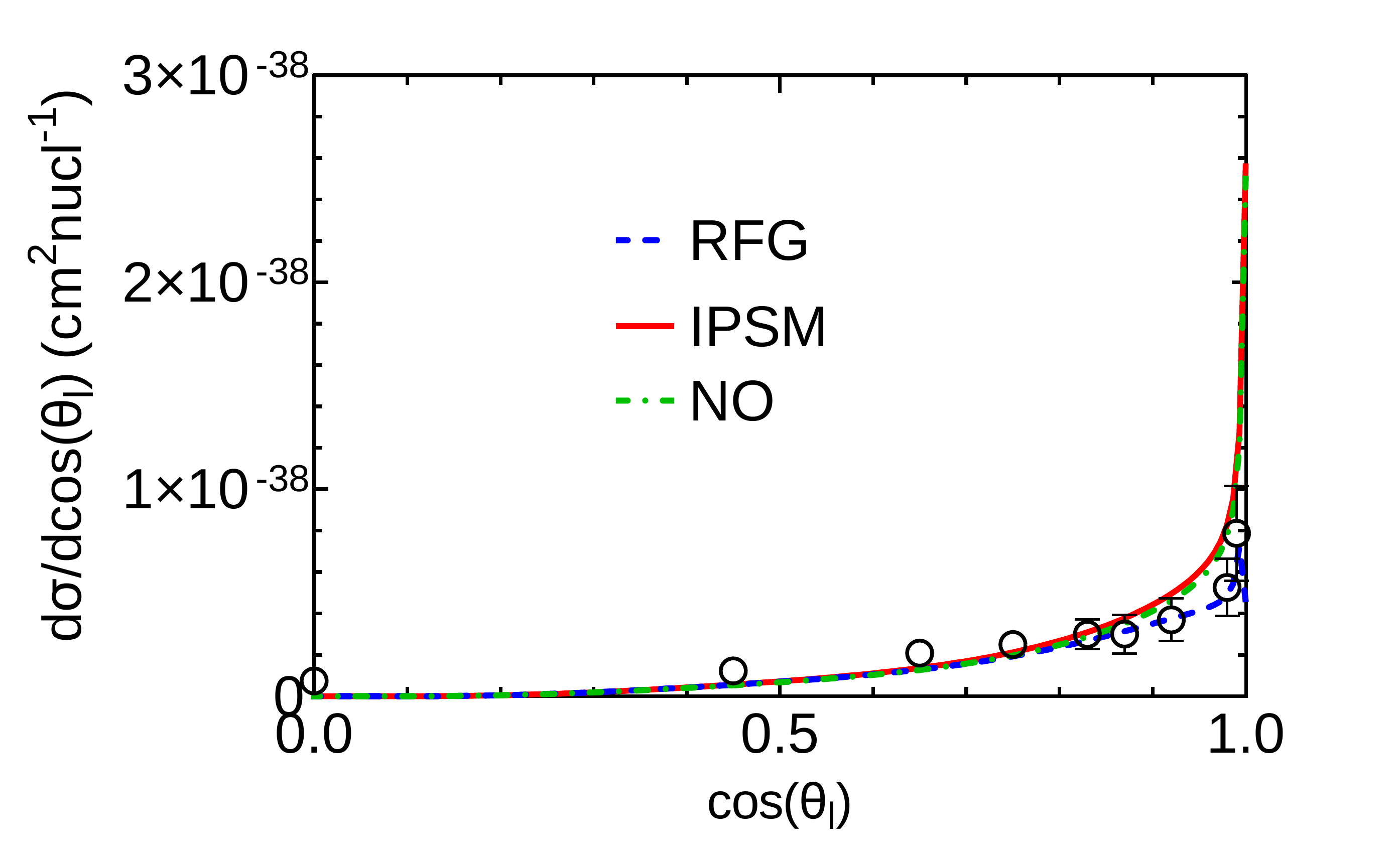}}
	\caption{\label{fig:T2K semi-inclusive integrated}T2K inclusive cross sections as function $\cos{\theta_l}$ obtained after integrating over all protons in the final state with momenta above 0.5 GeV (top panel) and with momenta below 0.5 GeV (bottom panel). Data taken from \cite{PhysRevD.98.032003}.}
\end{figure}

In Fig.~\ref{fig:T2K semi-inclusive integrated} we present the single differential inclusive cross section $d\sigma/d\cos\theta_l$ as a function of $\cos{\theta_l}$ when the integral over momenta of the ejected nucleon $p_N<0.5$ GeV  is performed (bottom panel) or restricting the analysis to $p_N > 0.5$ GeV (top panel). Comparing the two results, we can see that the difference between the RFG and other two models for $\theta_l$ close to zero observed 
for $p_N<0.5$ GeV is not present for $p_N>0.5$ GeV. This is a result of imposing a minimum value of the final proton momentum which is somehow equivalent to include Pauli blocking effects, because the orthogonality problem of the two shell models is hidden in this case by the experimental constraint on the final proton momentum.

\begin{figure*}[!htbp]
	\centering
	\includegraphics[width=\textwidth,height=0.85\paperheight]{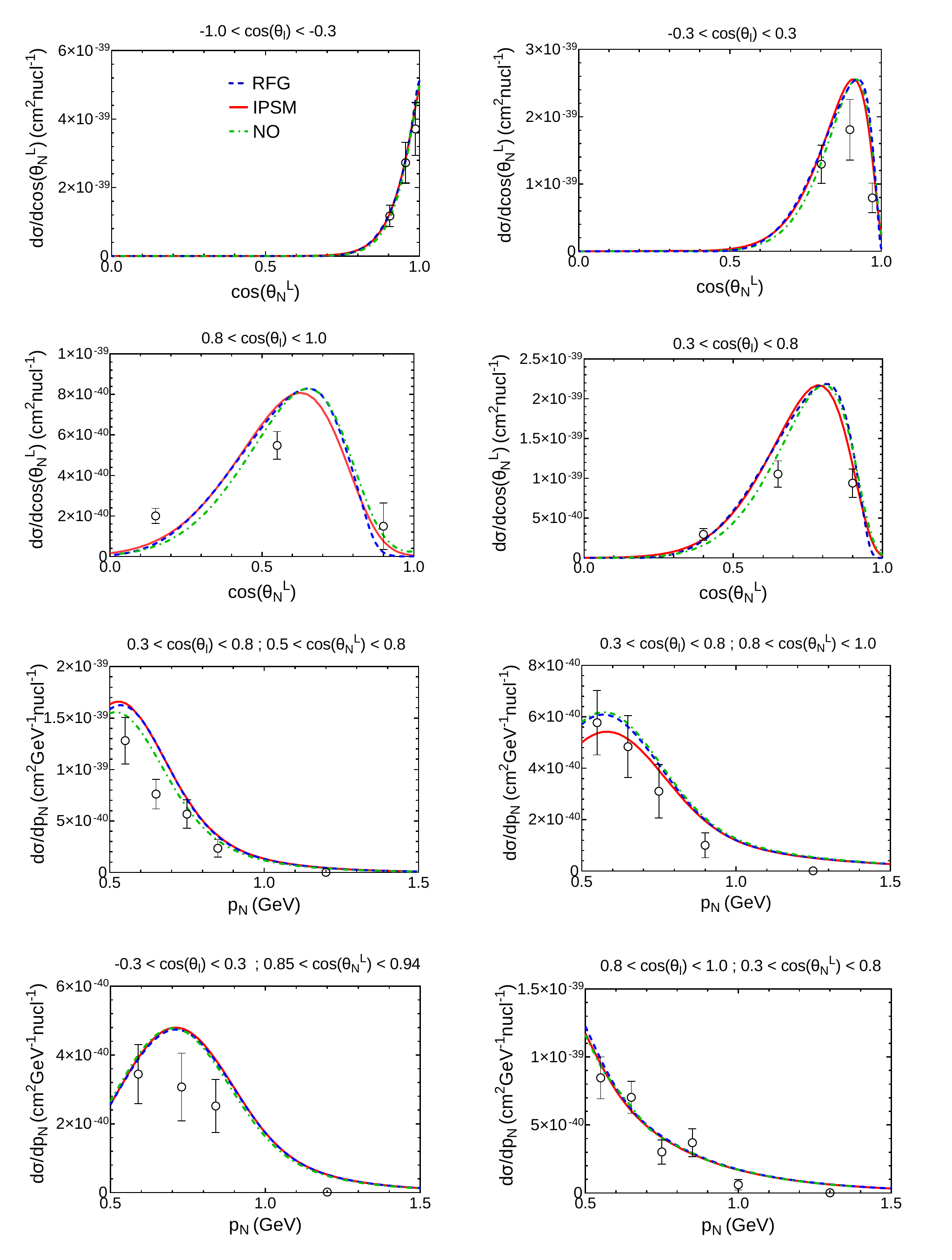}
	\caption{\label{fig:T2K semi-inclusive} The T2K CC0$\pi$1p single differential $\nu_\mu-^{12}$C cross sections as function of the NV. Data taken from \cite{PhysRevD.98.032003}.}
\end{figure*}

Moving now to the CC$0\pi1p$ data, in Fig.~\ref{fig:T2K semi-inclusive} we present single differential semi-inclusive $\nu_\mu - ^{12}$C cross sections with respect to the NV for the three nuclear models compared with available T2K data. Two different kinds of cross sections are considered: in the two top rows cross sections are presented as functions of $\cos{\theta_N^L}$ in bins of $\cos{\theta_l}$ and in the two bottom rows  as functions of $p_N$ in bins of $\cos{\theta_l}$ and $\cos{\theta_N^L}$. In all cases $p_N$ is larger than 0.5 GeV (see Table \ref{table:T2K constrains}). As shown, the uncertainty connected with the nuclear model is, in general, small and comparable with the one obtained for inclusive cross sections, taking into account that the experimental constraints imposed mask the orthogonality problem of the two shell models in the low-$q$ area.

In general, the interpretation of the discrepancies and agreements between
our results and the data shown in Figs.~\ref{fig:T2K inclusive}, \ref{fig:T2K semi-inclusive integrated} and \ref{fig:T2K semi-inclusive} is not straightforward since the measured cross sections are affected by multiple initial and final nuclear state effects which cannot be easily separated in the momentum and angular kinematic distributions. Note that the theoretical results presented  here only include the quasielastic regime and are based on the PWIA, neglecting FSI and 2p2h contributions which will be implemented in future work. A hint on the effects of these corrections is offered by simulations performed using the NEUT generator shown in \cite{PhysRevD.98.032003}, where the different effects of adding FSI and 2p2h as included in this generator are displayed. From this analysis it is shown that FSI increase the events without any proton with momentum above 0.5 GeV, thus enlarging the cross sections shown in Fig.~\ref{fig:T2K inclusive} and Fig.~\ref{fig:T2K semi-inclusive integrated} (bottom panel), and decreasing the cross sections shown in Fig.~\ref{fig:T2K semi-inclusive} and Fig.~\ref{fig:T2K semi-inclusive integrated} (top panel). The second (2p2h), according to the NEUT simulation, affect equally inclusive and semi-inclusive events by increasing the cross sections but, as it also happens for FSI, the effects for different bins can be very different. In our model, the complexity of the calculations make it difficult to predict how the cross sections will be modified by FSI and/or 2p2h before detailed calculations with strong relativistic scalar and vector potentials in the final state are performed (work in progress).

\begin{figure}[!htbp]
	\captionsetup[subfigure]{labelformat=empty} 
	\centering
	\subfloat[]{{\includegraphics[width=0.49\textwidth]{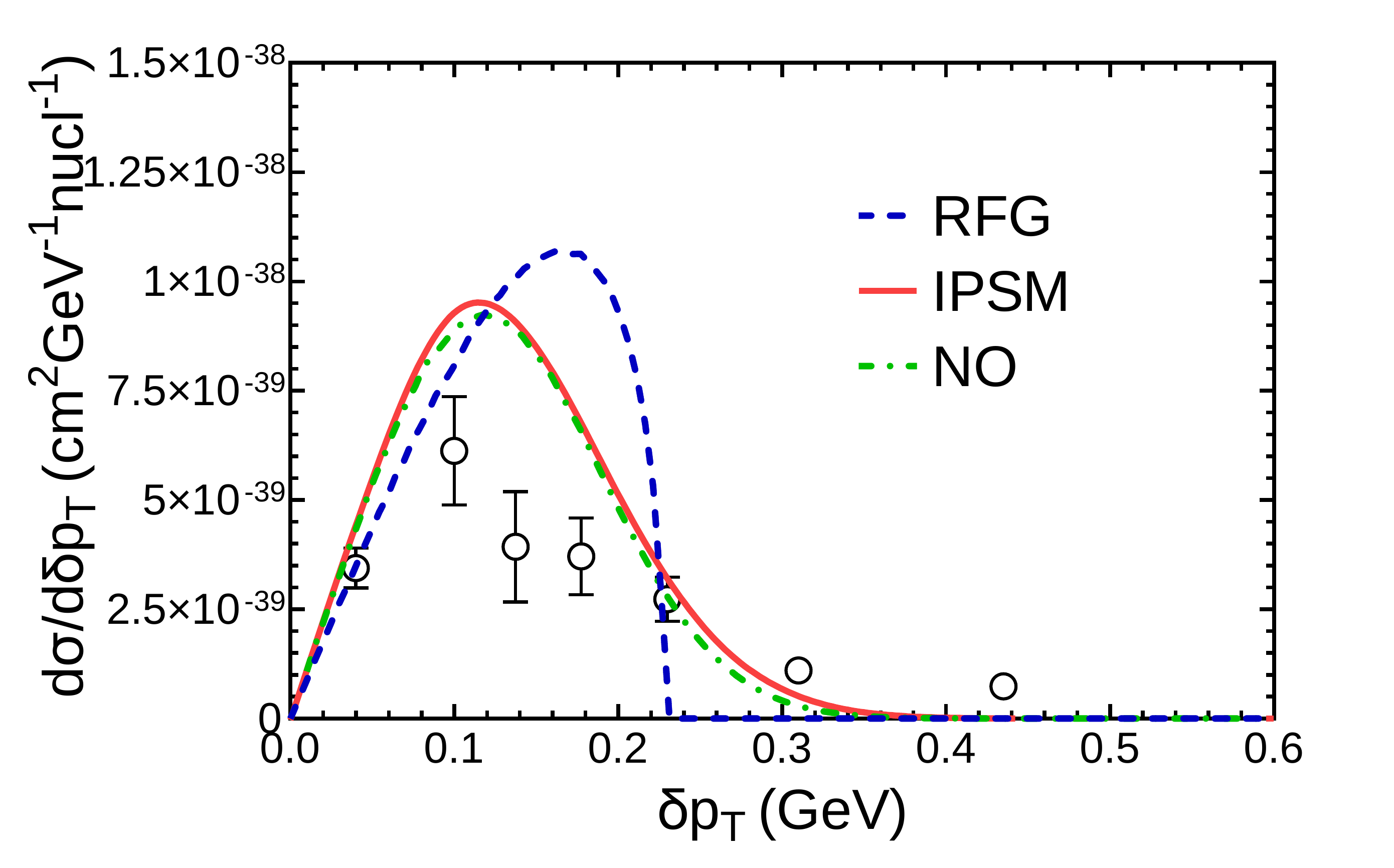}}}%
	\quad
	\subfloat[]{{\includegraphics[width=0.49\textwidth]{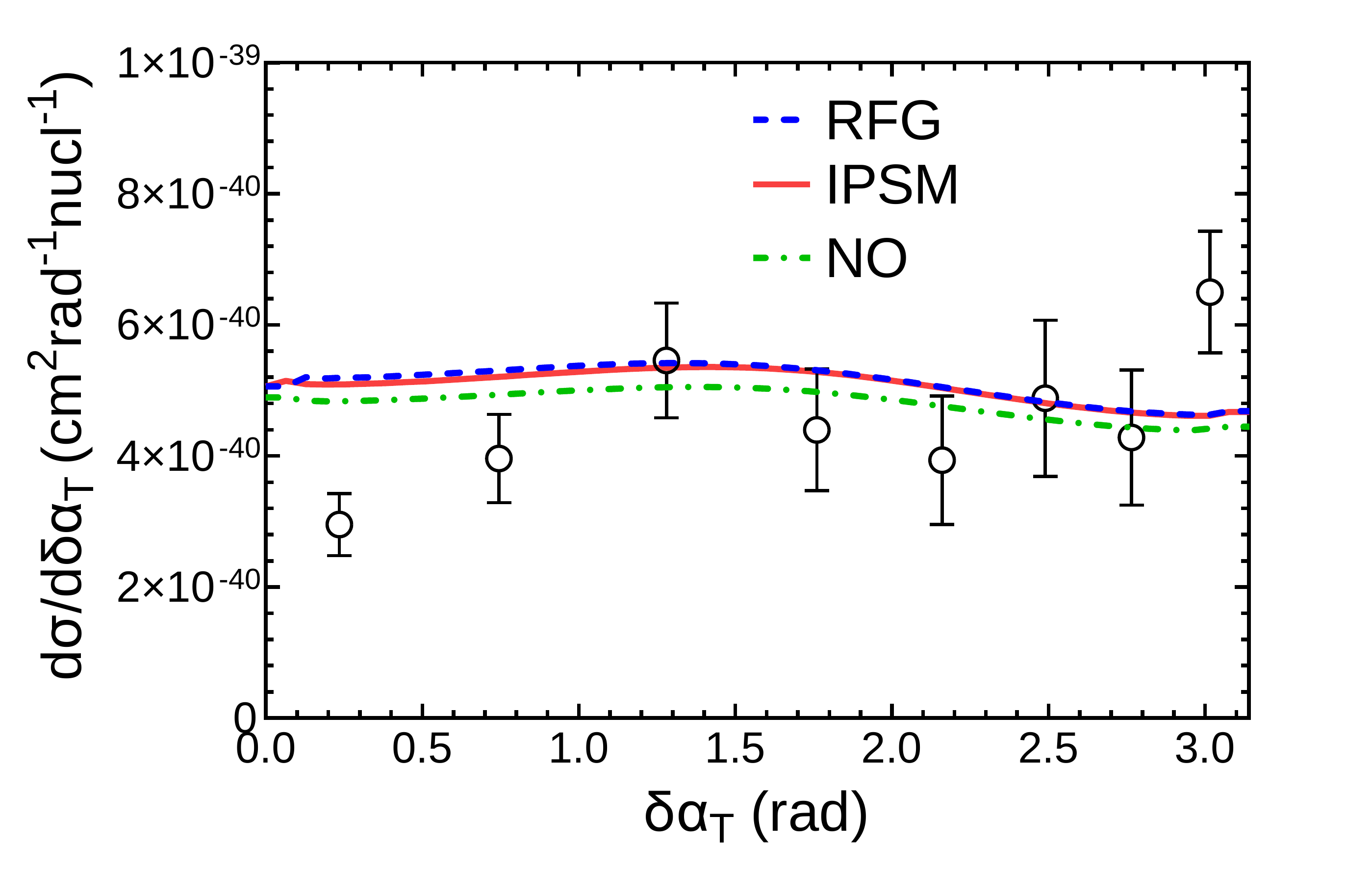}}}%
	\quad
	\subfloat[]{{\includegraphics[width=0.49\textwidth]{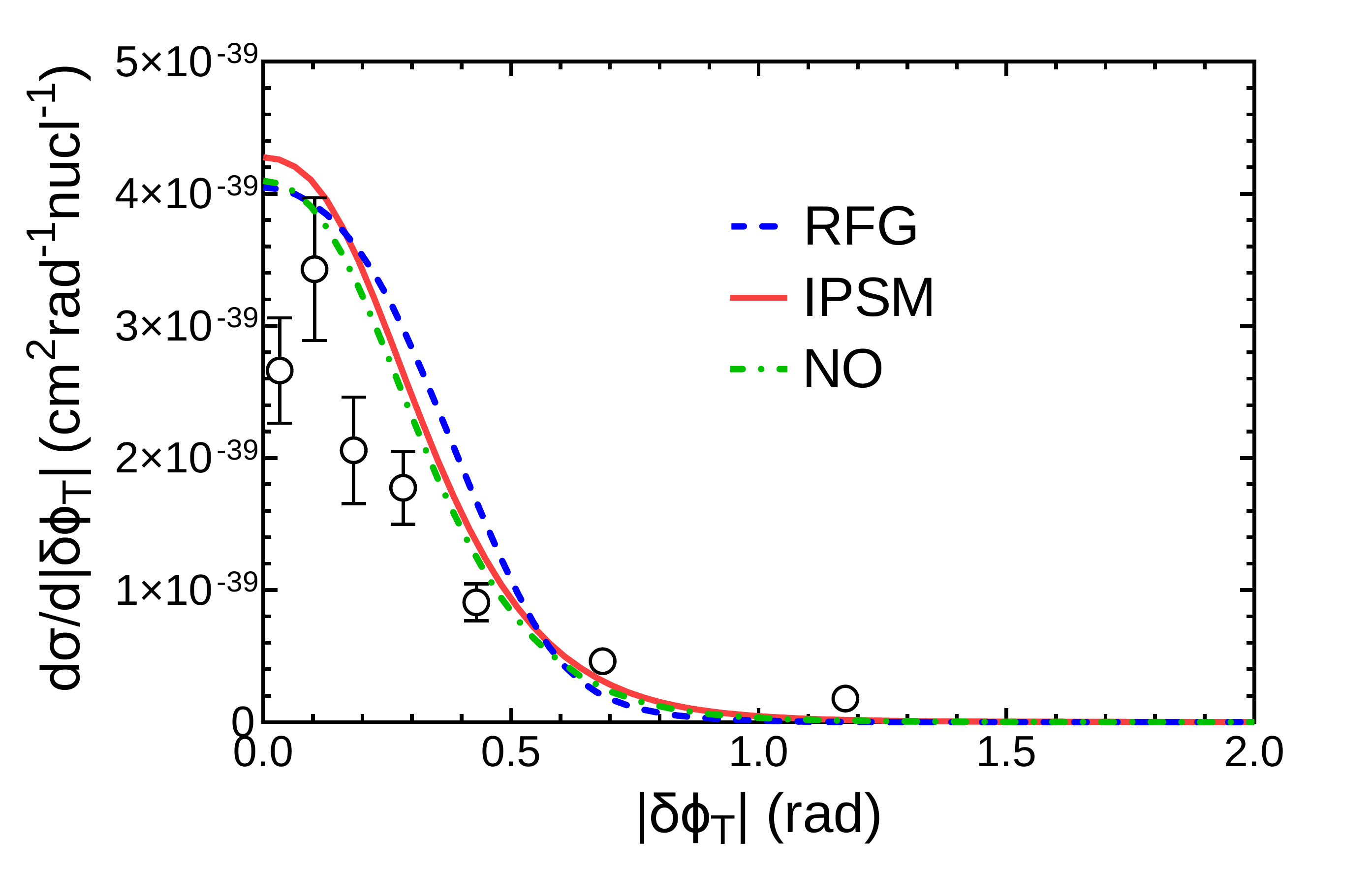}} }%
	\caption{\label{fig:T2K semi-inclusive STV} The T2K CC0$\pi$1p single differential $\nu_\mu-^{12}$C cross sections as function of the transverse kinematic imbalances $\delta p_T$, $\delta\alpha_T$ and $\left|\delta\phi_T\right|$ for the RFG model (blue dashed), IPSM (red solid) and NO (green dot-dashed). Data taken from \cite{PhysRevD.98.032003}.}
\end{figure}

Next we analyze the data in terms of the transverse kinematic imbalances defined in \eq{stv definition}. In Fig.~\ref{fig:T2K semi-inclusive STV} we show the semi-inclusive cross sections as function of TKI compared with the T2K data. As explained in Sec.~\ref{subsec:stv variables} this set of variables is designed to minimize effects associated to the neutrino energy. 
Only for $\delta\phi_T$ results are shown to depend more strongly on the kinematics of the incoming neutrino \cite{doi:10.7566/JPSCP.12.010032}. In the absence of FSI, the momentum imbalance is generated entirely by the description of the initial nuclear state dynamics. In this approximation, $\delta p_T$ is a direct measurement of the transverse component of the bound nucleon momentum distribution. As it can be seen in the top panel of Fig.~\ref{fig:T2K semi-inclusive STV}, the RFG cross section $d\sigma/d\delta p_T$ differs strongly from the other two, not only in magnitude and position of the maximum, but also in the fact that the RFG distribution vanishes for $\delta p_T>k_F$ as consequence of the Fermi condition. Although the position of the peak looks correct for the IPSM and NO models, the corresponding results overestimate the data in the low $\delta p_T$ area (below the Fermi momentum located around $0.23$ GeV for $^{12}$C) and underestimate the data for high $\delta p_T$, indicating that effects beyond PWIA might be essential to describe correctly the data. In the middle panel of Fig.~\ref{fig:T2K semi-inclusive STV} we show the cross sections as function of  $\delta\alpha_T$. In this case the NO prediction is a bit smaller than the RFG and IPSM ones, but all models exhibit a flat distribution as it was expected due to the PWIA and the isotropy of the associated momentum distributions. As it happened with $\delta p_T$, for $\delta\alpha_T$ we also see that PWIA is inadequate to describe the data. The inclusion of FSI and 2p2h contributions will likely improve the agreement with data. Finally, we also present the $\left|\delta\phi_T\right|$ distributions in the bottom panel of Fig.~\ref{fig:T2K semi-inclusive STV}, showing little discrepancies between the different models except for slightly smaller values of the cross sections around $\left|\delta\phi_T\right|=0$ for the RFG and NO models and a wider distribution for RFG compared with the other two models. As it happened with the other TKI, the PWIA does not give a good quantitative description of the data, although in this case it reproduces correctly the shape of the cross section.

\begin{figure*}[!htbp]
	\centering
	\includegraphics[width=\textwidth,height=0.80\paperheight]{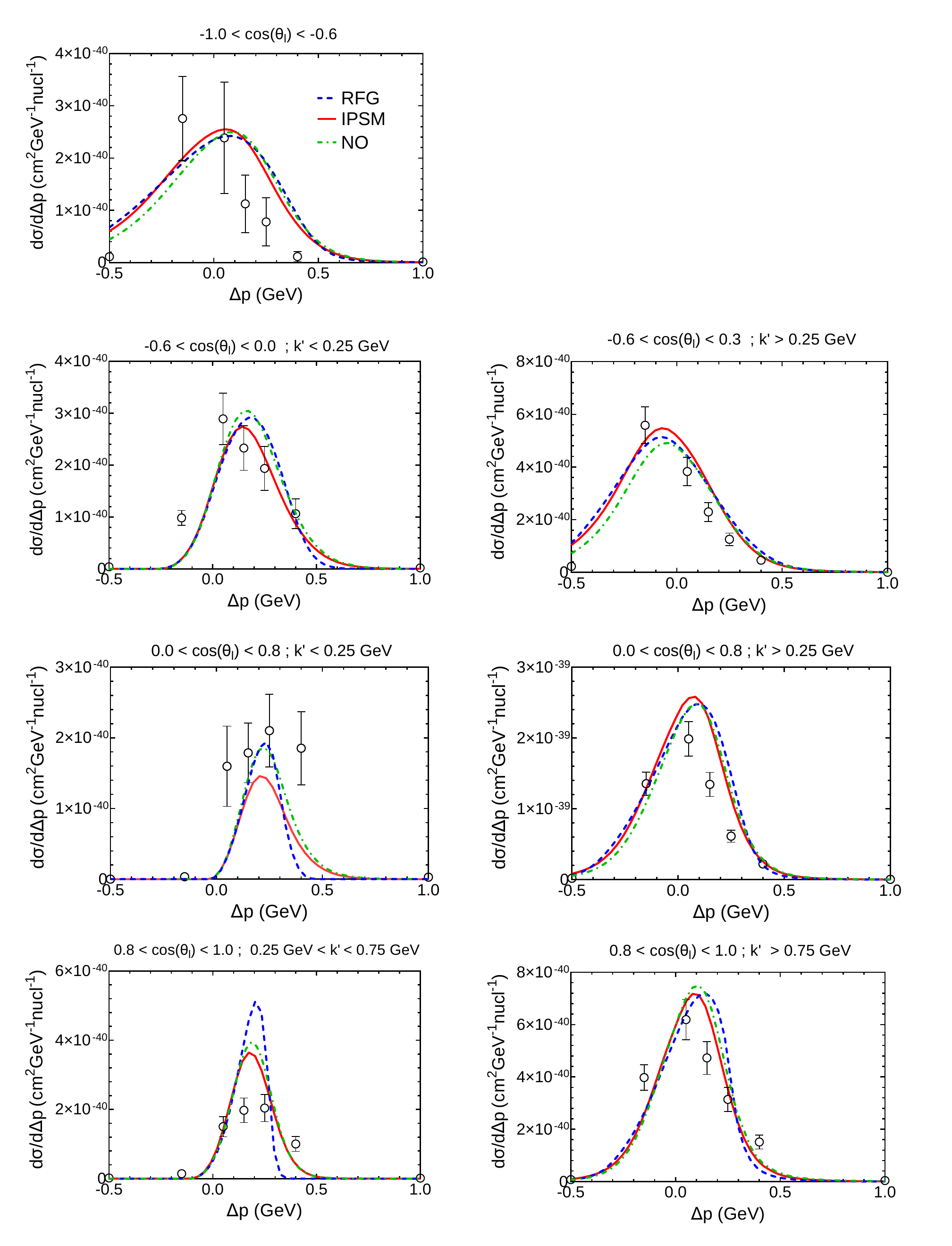}
	\caption{\label{fig:IV deltap}T2K CC0$\pi$1p single differential $\nu_\mu-^{12}$C cross section as function of the IV $\Delta p$ in different muon kinematic bins with constrains of the proton kinematics given in Table~\ref{table:T2K constrains} for the RFG (blue dashed), IPSM (red solid) and NO (green dot-dashed) nuclear models. Data taken from \cite{PhysRevD.98.032003}.}
\end{figure*}

\begin{figure*}[!htbp]
	\centering
	\includegraphics[width=\textwidth,height=0.80\paperheight]{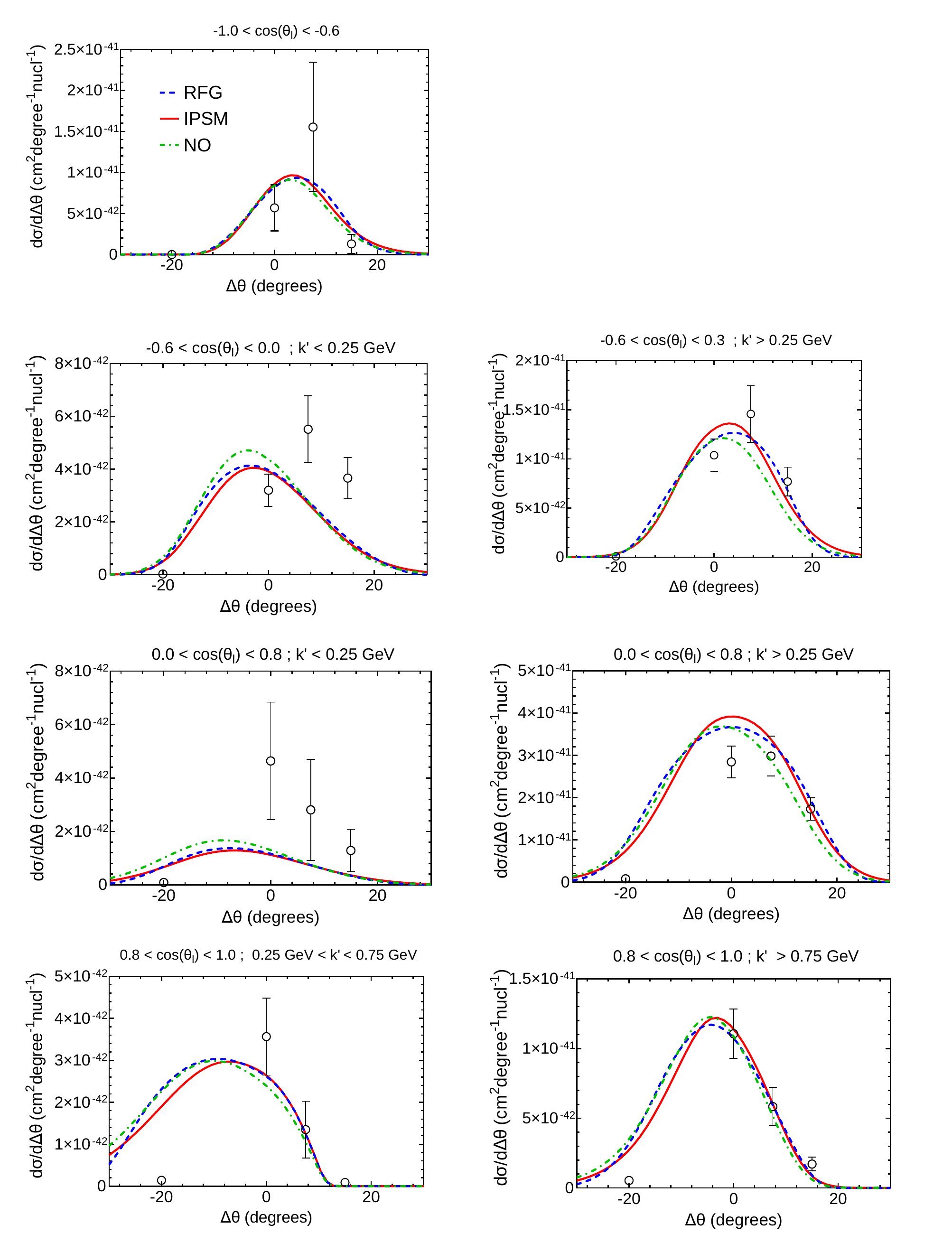}
	\caption{\label{fig:IV deltatheta}The T2K CC0$\pi$1p single differential $\nu_\mu-^{12}$C cross section as function of the IV $\Delta\theta$ in different muon kinematic bins with constrains of the proton kinematics given in Table~\ref{table:T2K constrains} for the RFG (blue dashed), IPSM (red solid) and NO (green dot-dashed) nuclear models. Data taken from \cite{PhysRevD.98.032003}.}
\end{figure*}

\begin{figure*}[!htbp]
	\centering
	\includegraphics[width=\textwidth,height=0.80\paperheight]{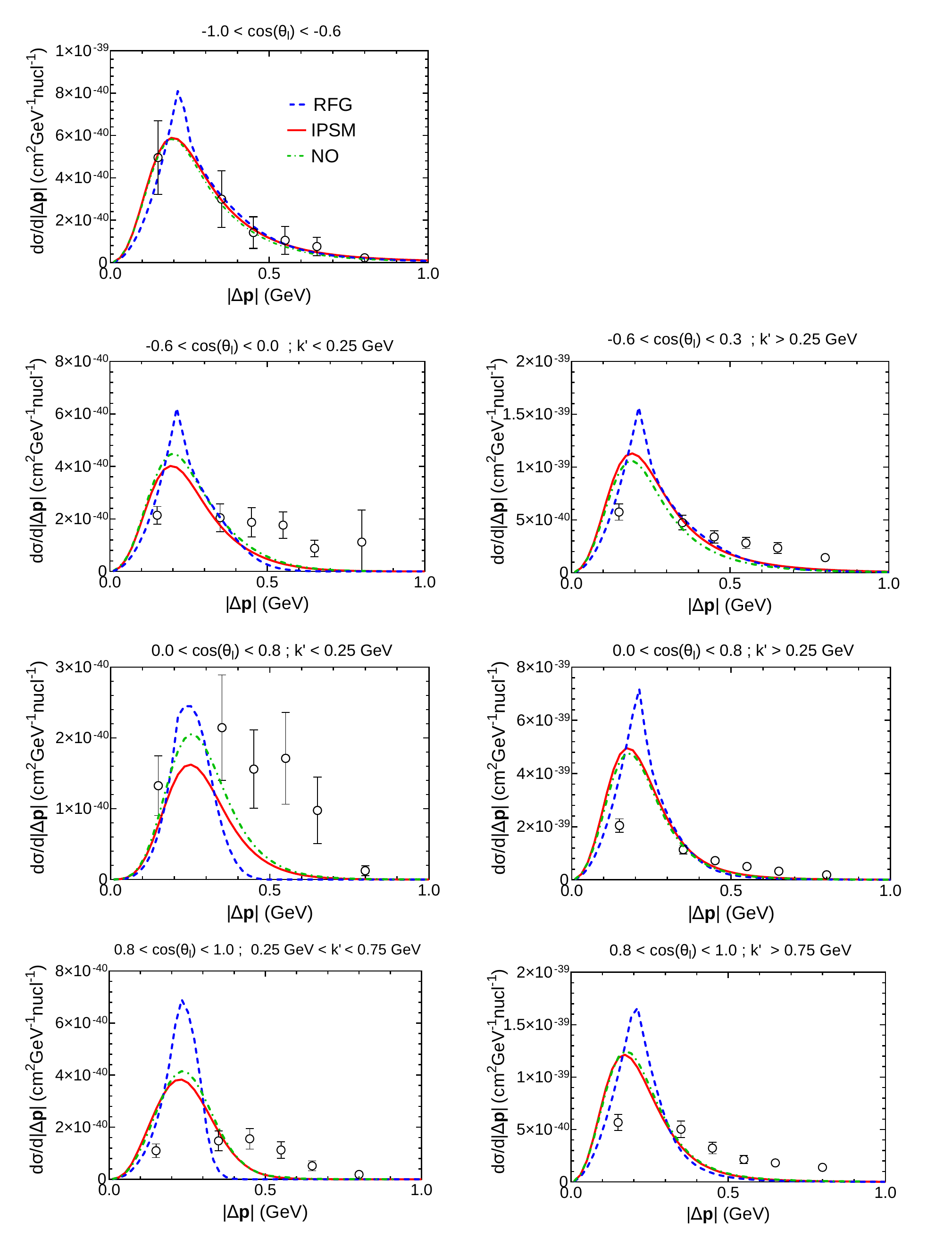}
	\caption{\label{fig:IV deltamodp}The T2K CC0$\pi$1p single differential $\nu_\mu-^{12}$C cross section as function of the IV $\left|\Delta\mathbf{p}\right|$ in different muon kinematic bins with constrains of the proton kinematics given in Table~\ref{table:T2K constrains} for the RFG (blue dashed), IPSM (red solid) and NO (green dot-dashed) nuclear models. Data taken from \cite{PhysRevD.98.032003}.}
\end{figure*}

To conclude with T2K analysis, in Figs.~\ref{fig:IV deltap}, \ref{fig:IV deltatheta} and \ref{fig:IV deltamodp} we present the semi-inclusive cross sections as function of the inferred variables $\Delta p$, $\Delta\theta$ and $\left|\Delta\mathbf{p}\right|$ in different regions distinguished by a specific bin of muon kinematics. As observed, the discrepancies between the different nuclear models depend on the particular kinematics considered being larger for low values of the muon kinematic bin, particularly in the case of $\Delta p$ (left panels in Fig.~\ref{fig:IV deltap}) and $\left|\Delta\mathbf{p}\right|$ (Fig.~\ref{fig:IV deltamodp}). In the latter it is remarkable the discrepancy between the RFG and the IPSM/NO models. A similar comment applies also to some specific kinematics in Fig.~\ref{fig:IV deltap}. On the contrary, the three models lead to rather similar results for the cross section as function of $\Delta\theta$ (Fig.~\ref{fig:IV deltatheta}). Our predictions are consistent with the simulations shown in \cite{PhysRevD.98.032003} for the kinematical regions where effects beyond the impulse approximation and FSI are expected to be minor. This is clearly illustrated in the results shown where some kinematics are very well described by the model, even being based on PWIA, whereas some other situations are completely off. The latter correspond to kinematics where the simulations show larger effects due to FSI and ingredients beyond the Impulse Approximation. However, this should be verified by the theoretical calculations, and the present study should be considered as a first step in providing a consistent comparison between data and a microscopic theoretical description of semi-inclusive neutrino-nucleus scattering reactions. 

\vspace{0.5cm}
\subsection{\label{subsec:minerva results}MINERVA}

Moving now to the comparison with the results presented by the MINER$\nu$A collaboration, in Table~\ref{table:Minerva constrains} we summarize the constraints in the kinematics of the final muon and proton applied to the data published in Refs.~\cite{PhysRevD.101.092001,PhysRevLett.121.022504}. In this case, we compare our results as function of NV and TKI with the latest available data from the MINER$\nu$A experiment.

\begin{table*}[htbp!]
	\centering
	\begin{tabular}{cccccccccccccccc} 
	\hline
		\toprule\toprule
		\Mark{MINER$\nu$A}              &  &   $k'$  &   & & $\cos{\theta_l}$ & &    &  $p_N$  &  && $\cos{\theta_N^L}$& & &$\phi_N^L$&\\\midrule
		All analyses               & & 1.5-10 GeV & &  &      $> 0.939$   &  &  &0.45-1.2 GeV& &  &    $> 0.342$    & &&-& \\ 
		\bottomrule\bottomrule
		\hline
	\end{tabular}
	\caption{\label{table:Minerva constrains}Final muon and proton phase-space restrictions applied to the CC0$\pi$ data shown by MINER$\nu$A collaboration in \cite{PhysRevD.101.092001,PhysRevLett.121.022504}. Note that, in contrast with what the T2K collaboration does in \cite{PhysRevD.98.032003}, for MINER$\nu$A the same restrictions are applied to the analyses of the single differential cross sections as function of any kind of variables, $i.e.$, NV and TKI.}
\end{table*}

\begin{figure*}[!htb]
	\centering
	\includegraphics[width=\textwidth]{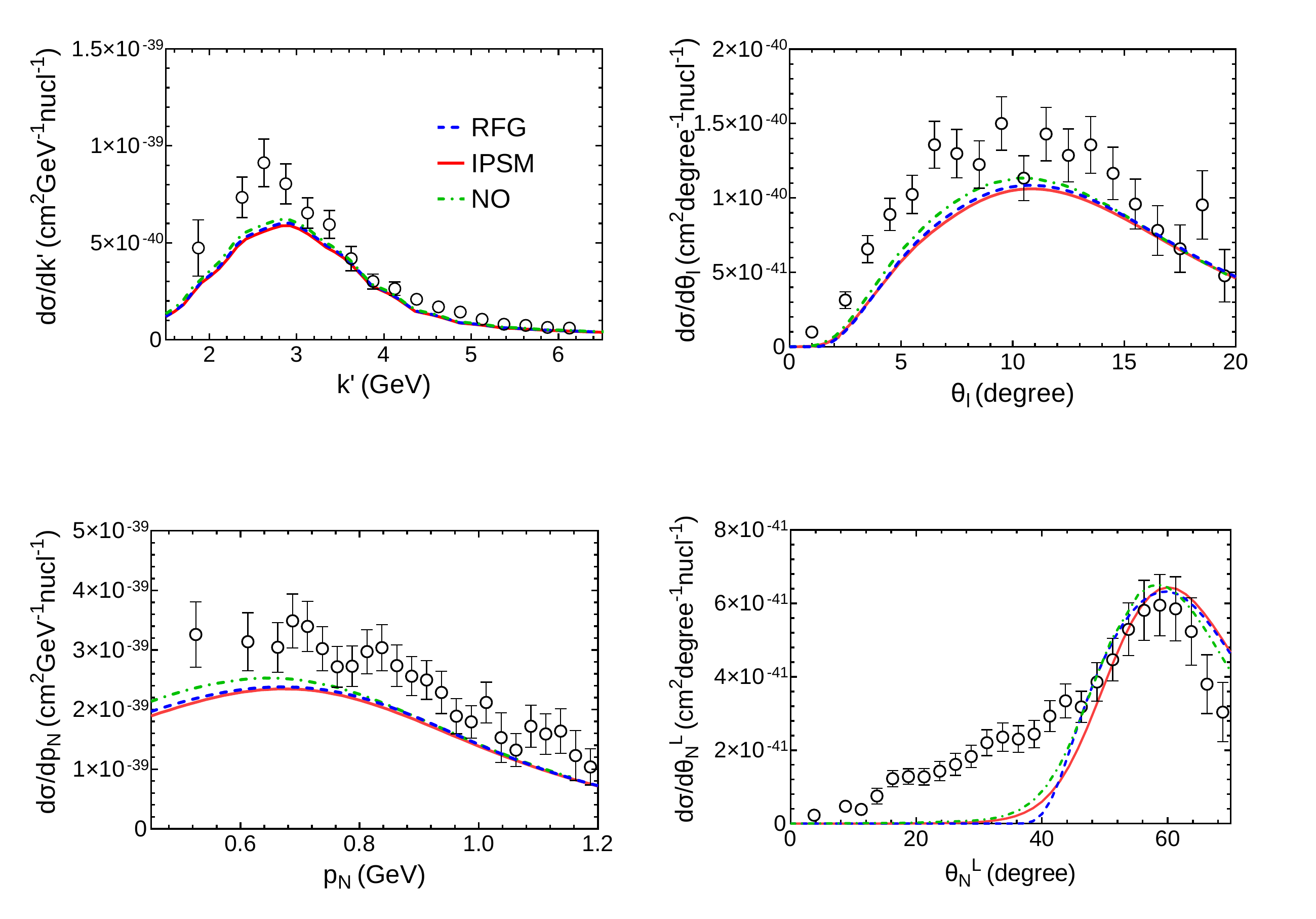}
	\caption{\label{fig:NV minerva} The MINER$\nu$A single differential inclusive $\nu_\mu-^{12}$C cross sections as function of the final muon momentum and scattering angle (top) and as function of the final proton momentum and polar angle (bottom) for the RFG (blue dashed), IPSM (red solid) and NO (green dot-dashed) nuclear models. The original paper with MINER$\nu$A data was \cite{PhysRevLett.121.022504} but the data shown here was taken from \cite{PhysRevD.101.092001} which corrected a mismodeling in GENIE's elastic FSI that affected the experimental data presented in the first paper.}
\end{figure*}

In Fig.~\ref{fig:NV minerva} we show the inclusive cross sections as function of the final muon momentum and scattering angle (top) and as function of the proton momentum and polar angle (bottom) for the three nuclear models considered. There is not any significant difference between the results in PWIA using the different nuclear models. In addition to this, our results reproduce very well the results generated by GENIE without FSI included in \cite{PhysRevLett.121.022504}, but systematically fall below the data. As pointed out in \cite{PhysRevLett.121.022504}, missing ingredients beyond the PWIA are necessary to describe correctly the experimental data, with a special mention to the area below $\theta_N^L \approx$ 40$^{\circ}$ where pion emission and re-absorption and 2p2h are, according to the GENIE simulation, the main ingredients that contribute to the large and long tail appreciated in the data.

\begin{figure}[!htbp]
	\captionsetup[subfigure]{labelformat=empty} 
	\centering
	\subfloat[]{{\includegraphics[width=0.49\textwidth]{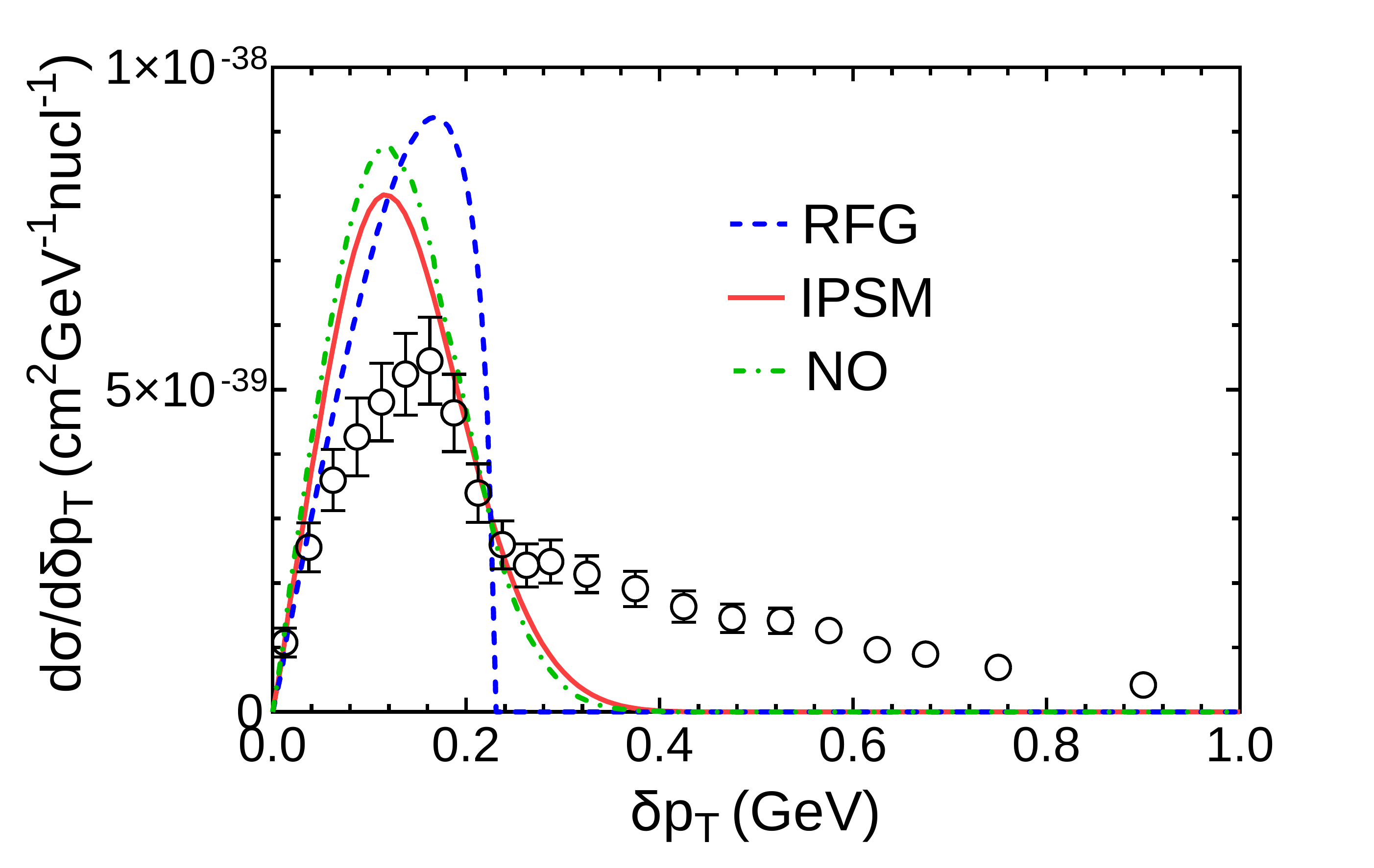}}}%
	\quad
	\subfloat[]{{\includegraphics[width=0.49\textwidth]{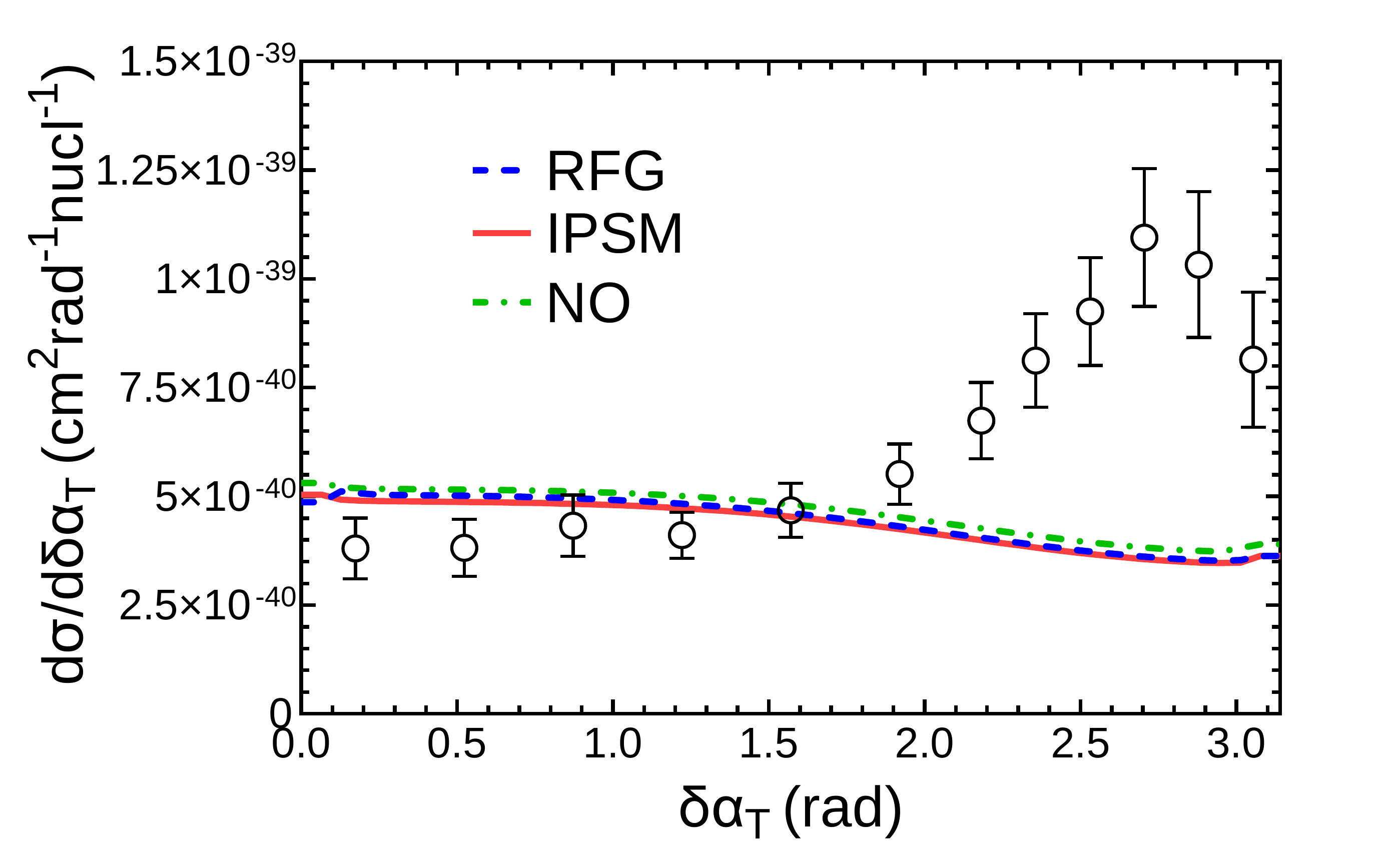}}}%
	\quad
	\subfloat[]{{\includegraphics[width=0.49\textwidth]{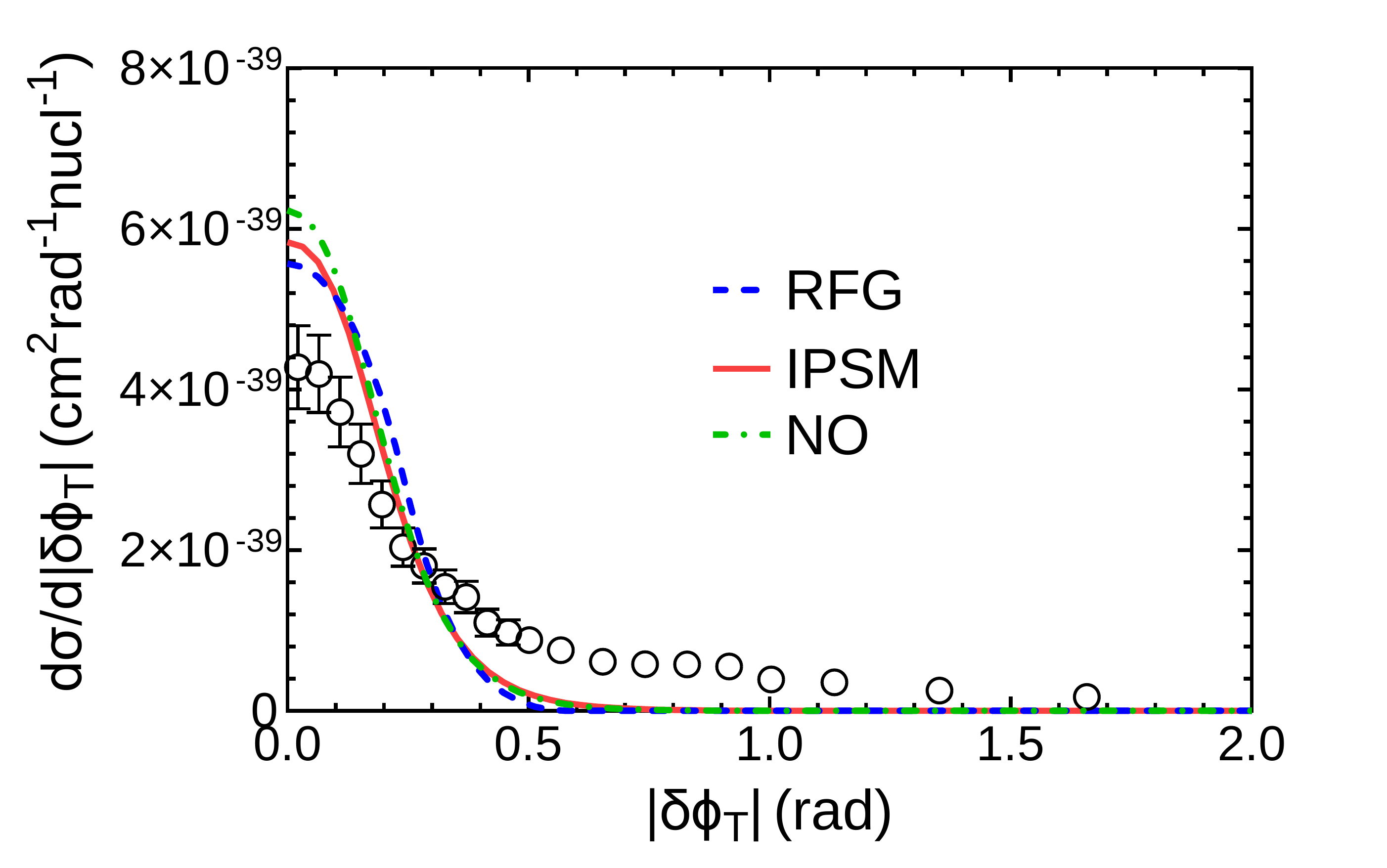}} }%
	\caption{\label{fig:STV minerva}The MINER$\nu$A CC0$\pi$1p single differential $\nu_\mu-^{12}$C cross sections as function of the transverse kinematic imbalances $\delta p_T$, $\delta\alpha_T$ and $\left|\delta\phi_T\right|$ for the RFG (blue dashed), IPSM (red solid) and NO (green dot-dashed) nuclear models. The original paper with MINER$\nu$A TKI data was \cite{PhysRevLett.121.022504} but the data shown here was taken from \cite{PhysRevD.101.092001} which corrected a mismodeling in GENIE's elastic FSI that affected the experimental data presented in the first paper.}
\end{figure}

In Fig.~\ref{fig:STV minerva} we present the differential cross sections as function of TKI compared with MINER$\nu$A data. As we already observed for T2K, the $\delta p_T$ distribution for the RFG model differs from the other two models in position of the peak and strength. This was expected because the momentum distribution of the RFG model is very different compared with the other two and $\delta p_T$ in the PWIA is the transverse component of the bound nucleon momentum distribution. Also, as observed for T2K, in this case the $\delta\alpha_T$ distribution is flat, as expected in the PWIA. Simulations using GENIE shown in \cite{PhysRevLett.121.022504} shed light on the differences observed between the data and the PWIA results, which are attributed to effects beyond the PWIA. Finally, note the theoretical prediction compared with data in the case of the cross section as function of $|\delta\phi_T|$. Whereas the PWIA overestimates data at low $|\delta\phi_T|$, the reverse occurs for increasing $|\delta\phi_T|$ where the tail shown by data is completely absent in the PWIA results.

\begin{figure}[!htbp]
	\captionsetup[subfigure]{labelformat=empty} 
	\centering
	{
	\subfloat[]{{\includegraphics[width=0.49\textwidth]{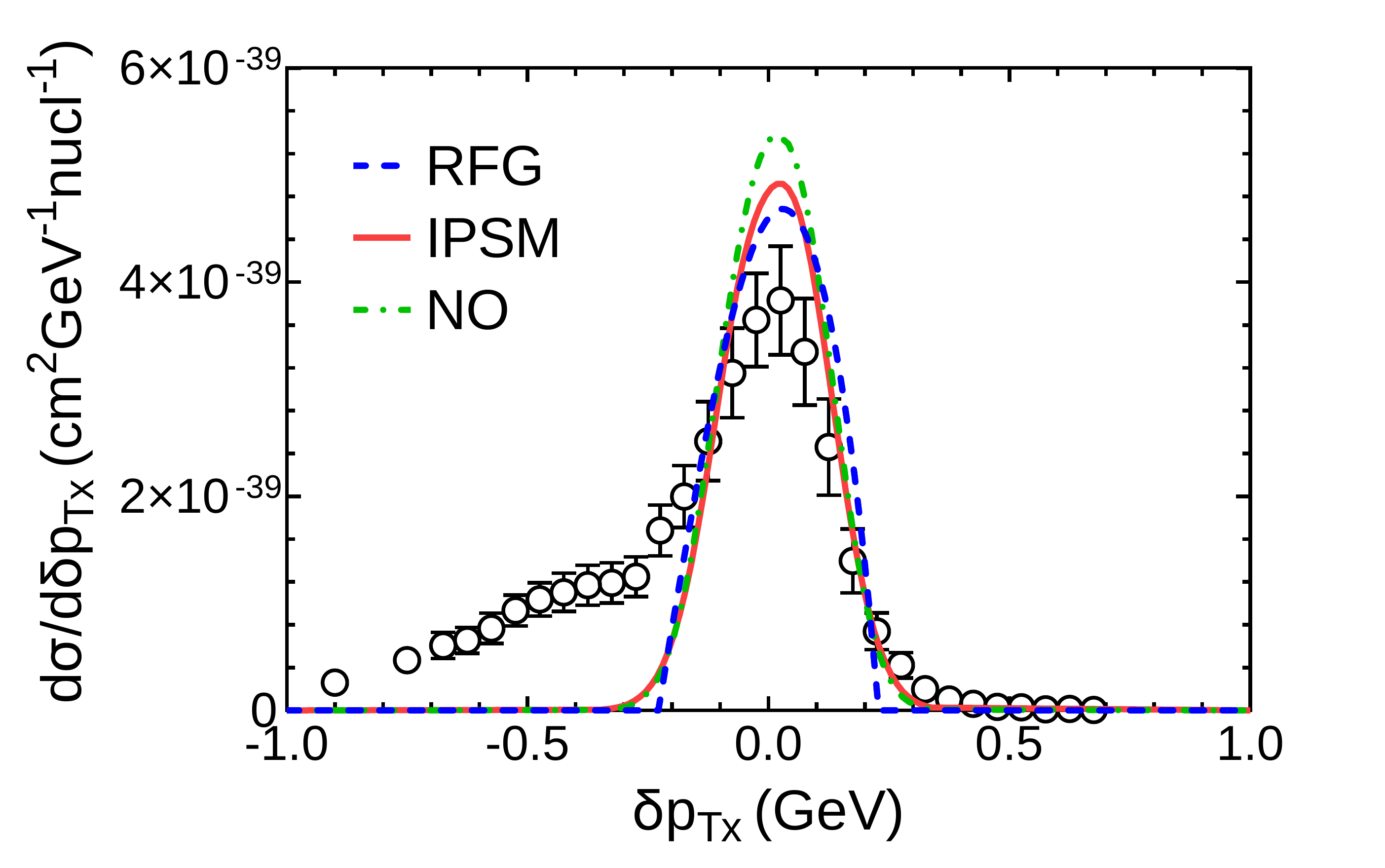}}}%
	\quad
	\subfloat[]{{\includegraphics[width=0.49\textwidth]{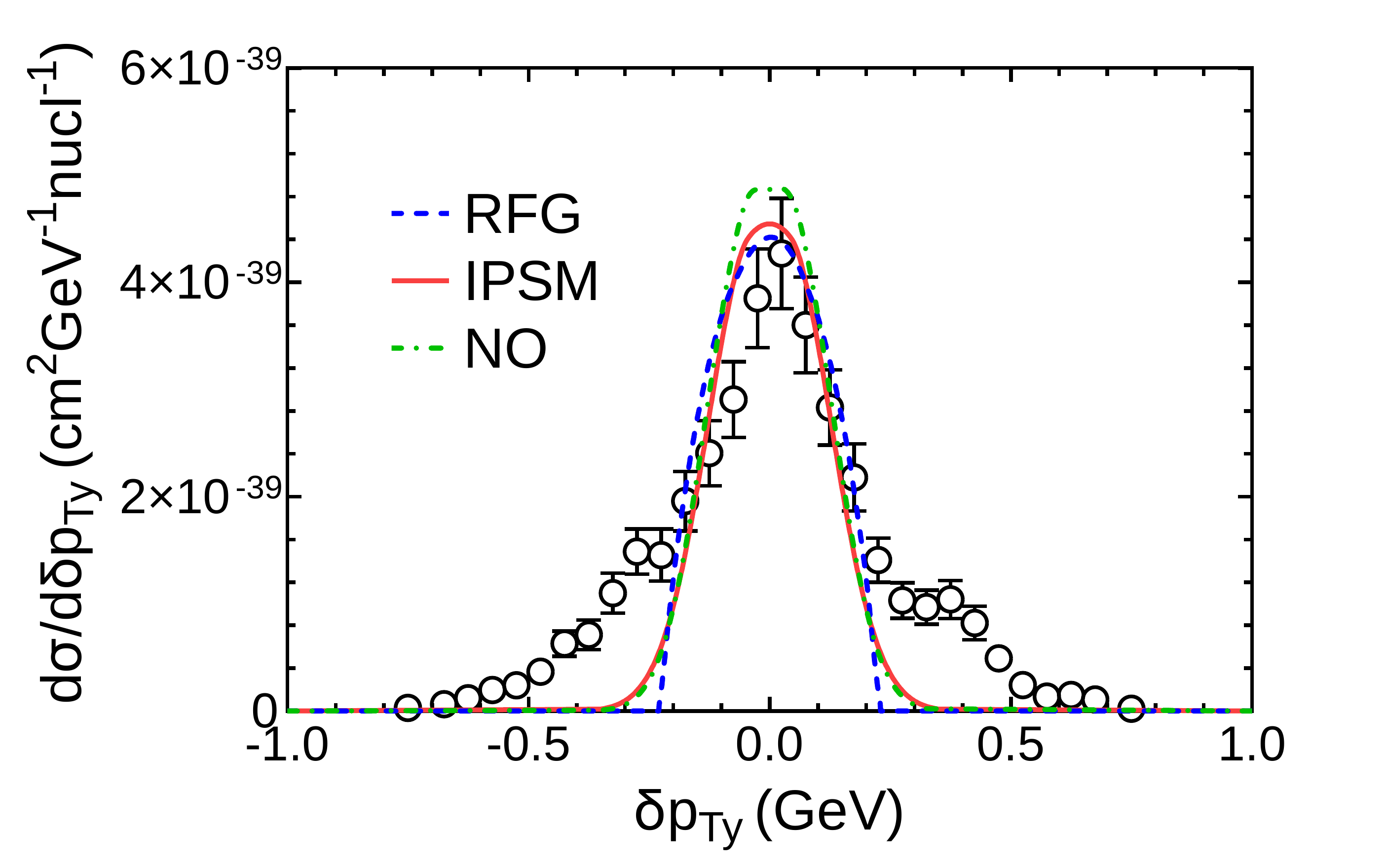}}}%
    }
    \caption{\label{fig:STV minerva projected} The MINER$\nu$A CC0$\pi$1p single differential $\nu_\mu-^{12}$C cross sections as function of the projections of $\delta p_T$ in the $x$ axis (left) and in the $y$ axis (right) for the RFG model (blue dashed), IPSM (red solid) and NO (green dot-dashed). Data taken from \cite{PhysRevD.101.092001}. Notice that the convention used in \cite{PhysRevD.101.092001} to define the $x$ and $y$ axis is the opposite to the convention used in this paper to define them, which is shown in Fig.~\ref{fig:stv}.}
\end{figure}

MINER$\nu$A also measured  cross sections versus the transverse components of the momentum imbalance. Following the definition given in \eq{stv definition}, the $x$ and $y$ components of the three-momentum imbalance are
\begin{align}
		\delta p_{Tx} &= k'\sin{\theta_l} + p_N\sin{\theta_N^L}\cos{\phi_N^L} \label{eq:dptx} \, , \\
		\delta p_{Ty} &= p_N\sin{\theta_N^L}\sin{\phi_N^L}\, . \label{eq:dpty}
\end{align}
In Fig.~\ref{fig:STV minerva projected} we compare our results for the differential cross sections as function of the components of the momentum imbalance with MINER$\nu$A data. If the interaction occurred on a free nucleon, then we would expect a delta-function at $\delta p_T = 0$ because the muon and proton final states would be perfectly balanced in that case, as required by momentum conservation. In the PWIA $\delta p_{Ty}$ is exactly the projection on the $y$-axis of the initial nucleon momentum. Because of the isotropy of the nucleon momentum distribution, the $\delta p_{Ty}$ distributions obtained in Fig.~\ref{fig:STV minerva projected} for all nuclear models are symmetrical around $\delta p_{Ty}$~=~0 and the width of the peaks are only result of the Fermi motion. On the other hand, $\delta p_{Tx}$ also depends on the final lepton kinematics, as seen in \eq{eq:dptx}. This produces a very slight shift of the peaks towards positive values of $\delta p_{Tx}$. Results in the PWIA are able to reproduce correctly the position of the peak but fail to match the long tails appreciated in the experimental data and also overestimate some of the data around the peak. Contributions beyond PWIA may reduce the discrepancies observed in the present analysis.

\subsection{\label{subsec:microboone results}MicroBooNE}

In the previous sections we have compared and analyzed experimental data of muon neutrinos on $^{12}$C published by the T2K and MINER$\nu$A collaborations. The MicroBooNE collaboration has also published CC0$\pi$1p flux-integrated differential cross sections of muon neutrinos on $^{40}$Ar \cite{microbooneA, microbooneB} as functions of the final particle momenta and angles. Next we present results compared with two sets of experimental measurements published by MicroBooNE. The kinematic constraints imposed in the two measurements are summarized in Table~\ref{table:Microboone constrains}.

\begin{table*}[!htbp]
    \centering
    \begin{tabular}{cccccccccccccccccccccc}
    \hline
    \toprule\toprule
        \Mark{MicroBooNE A}              &  &   $k'$  &   & & $\cos{\theta_l}$ & &    &  $p_N$  &  && $\cos{\theta_N^L}$&& &$\phi_N^L$& & & $\Delta\theta_{l,N}$ &&&$\delta p_T$&\\\midrule
    All analyses               & & $> 0.1$ GeV & &  &      -   &  &  &0.3-1.2 GeV& &  &    -    & &&- & & & -&&&-&\\\midrule
    \Mark{MicroBooNE B}              &  &    &   & &  & &    &    &  && &\\\midrule
    All analyses               & & 0.1-1.5 GeV & &  &      $-0.65 < \cos{\theta_l} < 0.95$   &  &  &0.3-1.0 GeV& &  &    $> 0.15$    & &&145-215$^\circ$ & & & 35-145$^\circ$& & &$\delta p_T < 0.35$ GeV&\  \\ \hline
    \bottomrule\bottomrule
\end{tabular}
\caption{\label{table:Microboone constrains}Muon and proton phase-space restrictions applied to the CC0$\pi$ data shown by MicroBooNE collaboration in \cite{microbooneA} (MicroBooNE A) and \cite{microbooneB} (MicroBooNE B). Here, $\Delta\theta_{l,N}$ is the opening angle defined as the angle between the muon and the proton and $\delta p_T$ the momentum imbalance in the transverse plane defined in \eq{stv definition}.}
\end{table*}

\begin{figure*}[!htb]
	\centering
	\includegraphics[width=\textwidth]{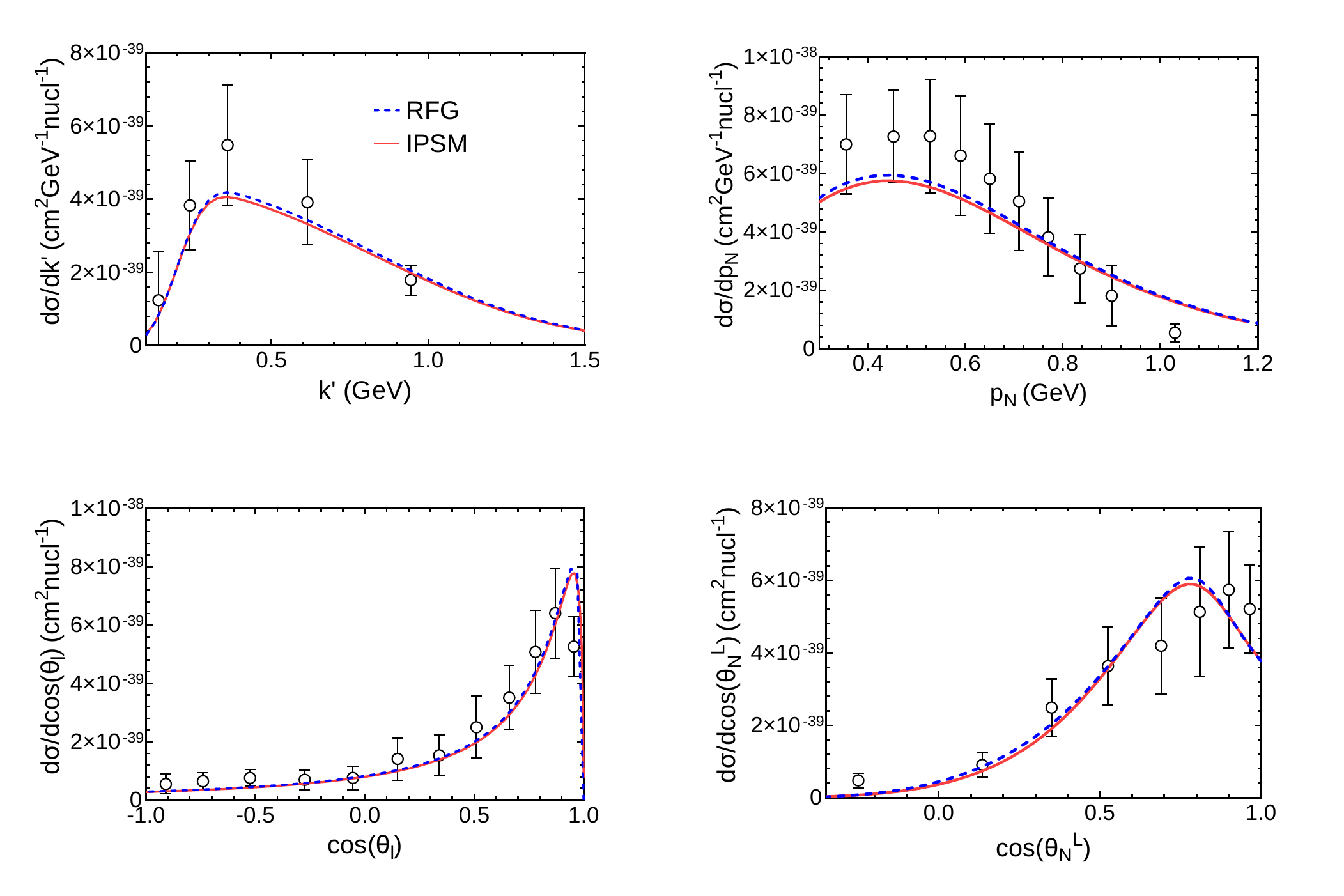}
	\caption{\label{fig:microboone A}The semi-inclusive CC0$\pi$1p single differential $\nu_\mu-^{40}$Ar cross sections as function of final muon and proton kinematics for MicroBooNE A data set. Experimental data was taken from \cite{microbooneA} and phase-space restrictions applied are summarized in Table~\ref{table:Microboone constrains}.}
\end{figure*}

In Fig.~\ref{fig:microboone A} we show single differential $\nu_\mu-^{40}$Ar cross sections as function of muon momentum and scattering angle (left panels) and as function of proton momentum and polar angle (right panels) for two nuclear models, RFG and IPSM, compared with data shown in \cite{microbooneA}. It is remarkable to point out how well the PWIA calculations reproduce the shape and magnitude of data. However, this does not contradict the important role that FSI and 2p2h contributions can play in the description of data for some kinematics according to GENIE simulations~\cite{microbooneA}.

\begin{figure*}[!htb]
    \centering
    \includegraphics[width=\textwidth]{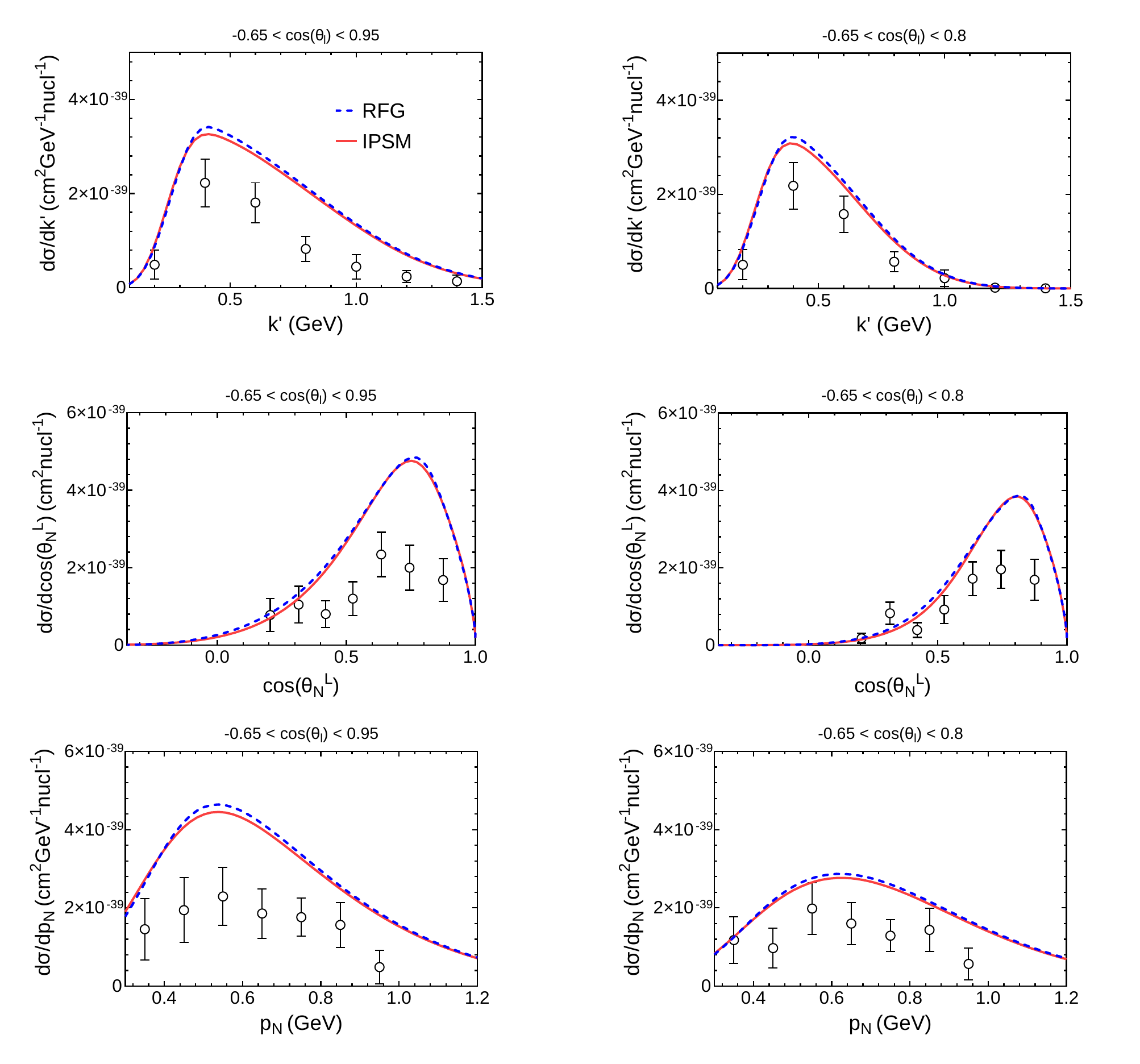}
    \caption{\label{fig:microboone B}The semi-inclusive CC0$\pi$1p single differential $\nu_\mu-^{40}$Ar cross sections as function of final muon and proton kinematics for MicroBooNE B data set. Experimental data was taken from \cite{microbooneB} and phase-space restrictions applied are summarized in Table~\ref{table:Microboone constrains}.}
\end{figure*}

To go deeper in the analysis of MicroBooNE data, in Fig.~\ref{fig:microboone B} we present again single differential $\nu_\mu-^{40}$Ar cross sections as function of muon momentum and proton momenta and polar angle but now compared with experimental data shown in \cite{microbooneB}. As explained there, the phase-space restrictions, which are summarized in Table~\ref{table:Microboone constrains}, were set to enhance CCQE interactions and to restrict the signal to the region where the detector response is well understood. Results are presented as function of the muon momentum, the proton momentum and polar angle, for two bins of $\cos{\theta_l}$, namely, $-0.65 < \cos{\theta_l} < 0.95$ (left panels) and $-0.65 < \cos{\theta_l} < 0.8$ (right panels). 

\begin{figure}[!htb]
	\centering
	\includegraphics[width=0.49\textwidth]{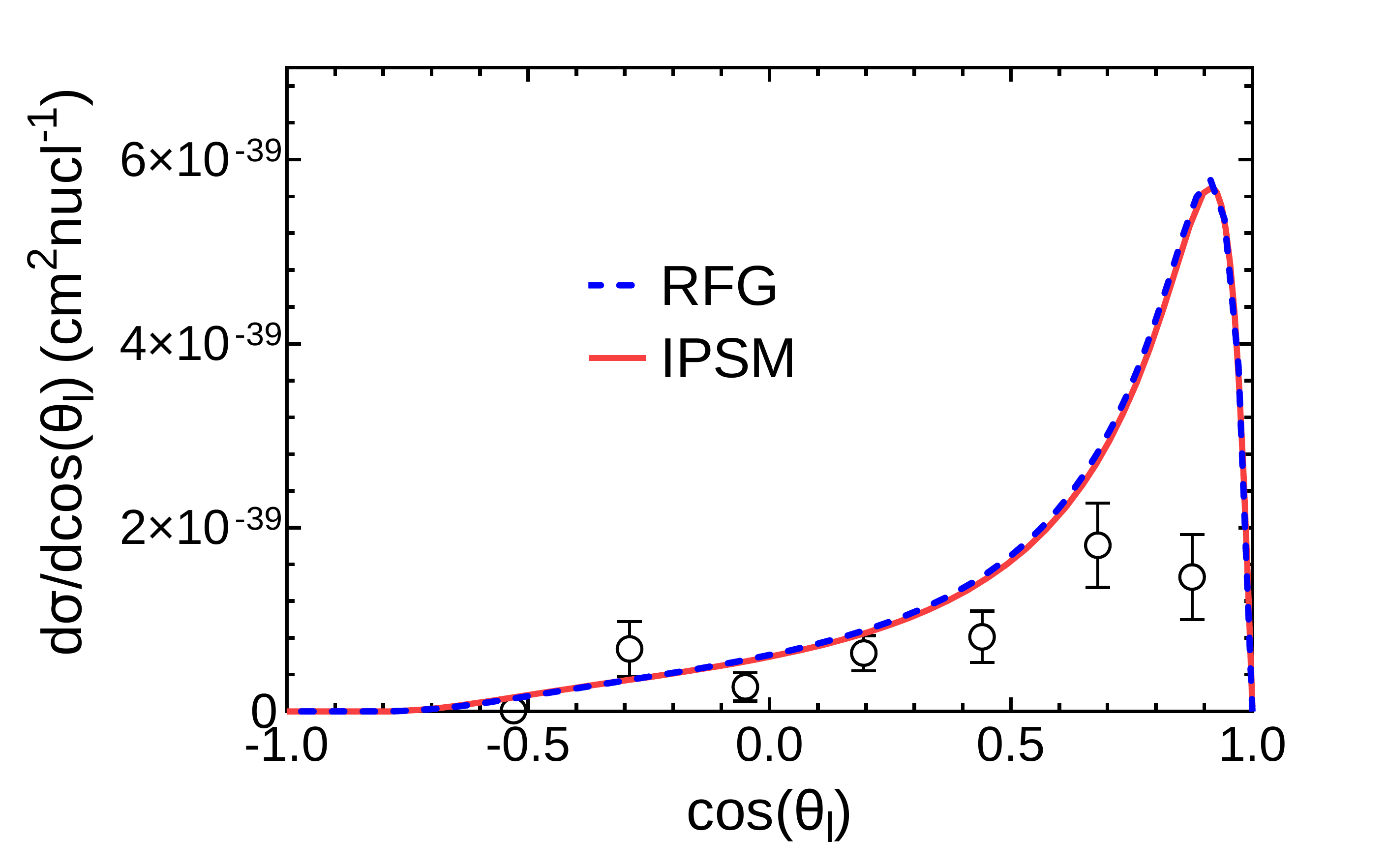}
	\caption{\label{fig:microboone B-costhetal}The semi-inclusive CC0$\pi$1p single differential $\nu_\mu-^{40}$Ar cross sections as function of $\cos{\theta_l}$ for MicroBooNE B data set. Experimental data was taken from \cite{microbooneB} and phase-space restrictions applied are summarized in Table~\ref{table:Microboone constrains}.}
\end{figure}

Although both the RFG and the IPSM model overestimate the two sets of data, the disagreement is larger in the left plots, where the muon scattering angle reaches smaller values. This is because in both models the cross section peaks in the area defined by $0.8 <\cos{\theta_l} < 0.95$, as shown in Fig.~\ref{fig:microboone B-costhetal}. This region corresponds to low values of $q$ and $\omega$, and excluding the contribution of this small-$\theta_l$ zone will produce a reduction of the cross section. This is clearly observed by comparing theoretical predictions in the left and right panels in Fig.~\ref{fig:microboone B}, and it is clearly in contrast with the data that only show a minor reduction when restricting the analysis to $\cos{\theta_l}<0.8$. This different behavior of theoretical predictions and data is consistent with the significant discrepancy shown in Fig.~\ref{fig:microboone B-costhetal} in the region of very small lepton scattering angles. A more careful study of this low-$\theta_l$ region is needed before more definite conclusions can be drawn.

\section{\label{sec:4}Conclusions}

In this paper we have analyzed all the available semi-inclusive CC0$\pi$ experimental data where a muon and at least one proton are detected in the final state from T2K, MINER$\nu$A and MicroBooNE neutrino collaborations. We have restricted ourselves to the plane wave impulse approximation, namely, ejected nucleons are described as plane waves and only one-body current operators are considered. To describe the nuclear dynamics, we have used three different nuclear models: the relativistic Fermi gas (RFG), the independent-particle shell model (IPSM) with fully relativistic Dirac wave function, and the natural orbitals (NO) shell model which includes NN correlations. 

Analytic expressions for the flux-averaged semi-inclusive cross sections for the three nuclear models are given as function of the momenta and angles of the muon and the proton detected in coincidence in the final state. Given that the experimental results were presented using different sets of kinematic variables, we have outlined the definition of each set, namely, transverse kinematic imbalances and inferred variables, and the explicit relationship between them and the final muon and proton momenta and angles measured in the laboratory system, denominated natural variables in this work.

Theoretical predictions for the cross sections as function of the muon and proton momenta and angles show very little dependence upon the specific nuclear model used in the PWIA. Specifically for T2K, differences appreciated in Fig.~\ref{fig:T2K inclusive} are related with the fact that IPSM and NO models produce much larger cross section than the data in the low momentum and energy transferred in the PWIA. Great caution must be taken when looking at semi-inclusive results presented in Fig.~\ref{fig:T2K semi-inclusive} because, although it might look like the theoretical predictions describe correctly some of the data, FSI and 2p2h contributions are not included. MINER$\nu$A results shown in Fig.~\ref{fig:NV minerva} fall below the data for almost any value of the kinematic variables used. This difference can be attributed to contributions beyond PWIA \cite{PhysRevLett.121.022504}. For MicroBooNE, results presented in Fig.~\ref{fig:microboone B} fall above the data even after excluding the contribution of $\cos{\theta_\mu} > 0.8$. We observe reductions of the cross sections similar to those obtained using Monte Carlo simulations when the events with $\cos{\theta_\mu} > 0.8$ are excluded, although the lack of effects beyond PWIA in our calculations does not allow us to find the root cause of the discrepancies between the results and the experimental data.

Concerning the use of variables linked with correlations between the muon and proton in the final state like the transverse kinematic imbalances, in the PWIA $\delta p_T$ is the projection in the transverse plane of the bound nucleon momentum, therefore different momentum distributions will produce different $\delta p_T$ distributions. As shown in Figs.~\ref{fig:T2K semi-inclusive STV}, \ref{fig:STV minerva}, and \ref{fig:STV minerva projected} the $\delta p_T$ distributions obtained using the RFG model are different from the ones obtained using the other two nuclear models. However, $\delta\alpha_T$ distributions obtained with the three nuclear models are similar in magnitude and shape, being all flat due to the isotropy of the momentum distribution and the absence of FSI in our calculations.

To conclude, semi-inclusive neutrino-nucleus reactions where a muon and one proton are detected in the final state can be used, with the right selection of experimental observables, to identify relevant nuclear effects related to both the initial state dynamics  and to final state interactions, as well as to two-particle-two-hole excitations and thus improve the reconstruction of the neutrino energy. The picture of the interaction drawn by the PWIA is an oversimplification of such complex processes, although it is a good starting point that highlights the importance of contributions beyond the PWIA necessary for the correct description of the available experimental data.
Work is in progress to extend the present analysis using more sophisticated descriptions of the final nucleon dynamics based on the Relativistic Mean Field (RMF). 

\begin{acknowledgments}
This work was partially supported by the Spanish Ministerio de Ciencia, Innovaci\'on y Universidades and ERDF (European Regional Development Fund) under contracts FIS2017-88410-P, by the Junta de Andaluc\'ia (grants No. FQM160 and SOMM17/6105/UGR) and by University of Tokyo ICRR's Inter-University Research Program FY2020 and FY2021. M.B.B. acknowledges support by the INFN under project Iniziativa Specifica and the University of Turin under Project BARM-RILO-20. J.M.F.P. acknowledges support from a fellowship from the Ministerio de Ciencia, Innovaci\'on y Universidades. Program FPI (Spain). G.D.M. acknowledges support from the European Unions Horizon 2020 research and innovation programme under the Marie Sklodowska-Curie grant agreement No. 839481.
\end{acknowledgments} 
  
\appendix

\section{\label{sec:app} Connection between TKI and NV}

In this appendix we deduce the Eqs.~\eqref{sdcs deltapt}, \eqref{sdcs deltaalphat}, and \eqref{sdcs deltaphit} which are the single differential cross sections with respect to the Transverse Kinematic Imbalances defined in \eq{stv definition}.

Starting with $\delta p_T$, we get 
\begin{align}\label{deltapt}
    \delta p_T^2 = \bigl( k'\sin{\theta_l} &+ p_N\sin{\theta_N^L}\cos{\phi_N^L} \bigr)^2 + \bigl( p_N\sin{\theta_N^L}\sin{\phi_N^L} \bigr)^2 \, ,
\end{align}
which can be written as a second degree equation for $p_N$ in the form $ap_N^2 + 2bp_N + c = 0$ with coefficients given by
\begin{align}\label{second degree deltapt}
    a &= \sin^2{\theta_N^L}, \nonumber \\
    b &= k'\sin{\theta_l}\sin{\theta_N^L}\cos{\phi_N^L}, \nonumber \\
    c &= k'^2\sin^2{\theta_l} - \delta p_T^2.
\end{align}
Changing $p_N \rightarrow \delta p_T$ and integrating the flux-averaged sixth differential cross section over all the other variables, we  obtain the single differential cross section as function of $\delta p_T$
\begin{widetext}    
    \begin{align}\label{sdcs deltapt2}
        \frac{d\sigma}{d\delta p_T} = 2\pi\int_{\theta_l^{min}}^{\theta_l^{max}}d\theta_l\sin{\theta_l}\int_{k'_{min}}^{k'_{max}}dk'\int_{\theta_N^{Lmin}}^{\theta_N^{Lmax}}d\theta_N^L\sin{\theta_N^L}\int_{\phi_N^{Lmin}}^{\phi_N^{Lmax}}d\phi_N^L
        \times\left < d^5\sigma \right > \times\mathcal{J}_p\times\Theta(p_N - p_N^{min})\Theta(p_N^{max}-p_N)
    \end{align}
\end{widetext}
with $\mathcal{J}_p$ the Jacobian at fixed $(\theta_N,\phi_N)$ expressed as 
\begin{align}\label{jacopt}
    \mathcal{J}_p =\biggl|\frac{\partial p_N}{\partial(\delta p_T)} \biggr|_{(\theta_N,\phi_N)} = \biggl|\frac{\delta p_T}{ap_N + b}\biggr| 
\end{align}
and $\left < d^5\sigma \right >$ the flux-averaged fifth-differential semi-inclusive cross section.
Note that the integrals in \eq{sdcs deltapt2} are performed between general minimum and maximum values. As can be seen in the results section, different kinematic constrains are imposed by the different neutrino collaborations that need to be taken into account when calculating theoretical predictions to compare with experimental data. Moreover, given a set of variables $(k', \theta_l, \delta p_T, \theta_N^L, \phi_N^L)$ the solution of the second degree equation for $p_N$ can give none, one or two valid solutions, understanding as a valid solution a real positive number that fulfills $p_N^{min} < p_N < p_N^{max}$.
In case of multiple valid solutions, their contributions to the cross section are summed.

Moving to the next variable, $\delta\alpha_T$, we find
\begin{align}\label{alphat}
    \cos{\delta\alpha_T} = \frac{-\bigl(k'\sin{\theta_l + p_N\sin{\theta_N^L}\cos{\phi_N^L}}\bigr)}{\delta p_T}
\end{align}
with $\delta p_T$ given in \eq{deltapt}. Squaring both sides of this equation and solving for $\theta_N^L$ we get
\begin{align}\label{sinthetanl}
    \sin{\theta_N^L} = \frac{k'\sin{\theta_l}\tan{\delta \alpha_T}}{p_N\bigl(\pm\sin{\phi_N^L} - \tan{\delta \alpha_T}\cos{\phi_N^L} \bigr)},
\end{align}
where $\pm$ arises after taking square root. Therefore, the single differential cross section can be expressed as
\begin{widetext}
    \begin{align}\label{sdcs deltaalphat2}
        \frac{d\sigma}{d\delta \alpha_T} = 2\pi\int_{\theta_l^{min}}^{\theta_l^{max}}d\theta_l\sin{\theta_l}\int_{k'_{min}}^{k'_{max}}dk'\int_{p_N^{min}}^{p_N^{max}}dp_N\int_{\phi_N^{Lmin}}^{\phi_N^{Lmax}}d\phi_N^L
        \left < d^5\sigma \right >\mathcal{J}_\alpha\sin{\theta_N^L}\Theta(\theta_N^L - \theta_N^{Lmin})\Theta(\theta_N^{Lmax}-\theta_N^L)
    \end{align} 
\end{widetext}
with $\mathcal{J}_\alpha$ the jacobian of the change $\theta_N^L \rightarrow \delta\alpha_T$ given by
\begin{align}\label{jacoalpha}
    \mathcal{J}_\alpha = \biggl| \frac{\partial \theta_N^L}{\partial \delta \alpha_T} \biggr| = \biggl| \frac{\bigl( 1 + \tan^2{\delta \alpha_T}\bigr)\tan{\theta_N^L}\sin{\phi_N^L}}{\tan{\delta \alpha_T}\bigl(\pm\sin{\phi_N^L} - \tan{\delta \alpha_T}\cos{\phi_N^L}\bigr)}  \biggr|.
\end{align}
For a fixed set of variables $(k', \theta_l, p_N, \delta\alpha_T, \phi_N^L)$ the value of $\theta_N^L$ is calculated following \eq{sinthetanl}. To be considered as a valid solution, it must satisfy \eq{alphat} and
$\theta_N^{Lmin} < \theta_N^L < \theta_N^{Lmax}$. 

Finally, $\delta\phi_T$ can easily be linked with $\phi_N^L$. Following the definition given in \eq{stv definition}, 
\begin{equation}
    \delta\phi_T = \phi_N^L - \pi \, .
\end{equation}
Henceforth, the single differential cross section as function of $\delta\phi_T$ is 
\begin{widetext}    
    \begin{align}
        \frac{d\sigma}{d\delta \phi_T} = 2\pi\int_{\theta_l^{min}}^{\theta_l^{max}}d\theta_l\sin{\theta_l}\int_{k'_{min}}^{k'_{max}}dk'\int_{p_N^{min}}^{p_N^{max}}dp_N\int_{\theta_N^{Lmin}}^{\theta_N^{Lmax}}d\theta_N^L\left < d^5\sigma \right >\sin{\theta_N^L}\, .
    \end{align}
\end{widetext}

\bibliography{references}

\begin{thebibliography}{29}%
\makeatletter
\providecommand \@ifxundefined [1]{%
 \@ifx{#1\undefined}
}%
\providecommand \@ifnum [1]{%
 \ifnum #1\expandafter \@firstoftwo
 \else \expandafter \@secondoftwo
 \fi
}%
\providecommand \@ifx [1]{%
 \ifx #1\expandafter \@firstoftwo
 \else \expandafter \@secondoftwo
 \fi
}%
\providecommand \natexlab [1]{#1}%
\providecommand \enquote  [1]{``#1''}%
\providecommand \bibnamefont  [1]{#1}%
\providecommand \bibfnamefont [1]{#1}%
\providecommand \citenamefont [1]{#1}%
\providecommand \href@noop [0]{\@secondoftwo}%
\providecommand \href [0]{\begingroup \@sanitize@url \@href}%
\providecommand \@href[1]{\@@startlink{#1}\@@href}%
\providecommand \@@href[1]{\endgroup#1\@@endlink}%
\providecommand \@sanitize@url [0]{\catcode `\\12\catcode `\$12\catcode
  `\&12\catcode `\#12\catcode `\^12\catcode `\_12\catcode `\%12\relax}%
\providecommand \@@startlink[1]{}%
\providecommand \@@endlink[0]{}%
\providecommand \url  [0]{\begingroup\@sanitize@url \@url }%
\providecommand \@url [1]{\endgroup\@href {#1}{\urlprefix }}%
\providecommand \urlprefix  [0]{URL }%
\providecommand \Eprint [0]{\href }%
\providecommand \doibase [0]{https://doi.org/}%
\providecommand \selectlanguage [0]{\@gobble}%
\providecommand \bibinfo  [0]{\@secondoftwo}%
\providecommand \bibfield  [0]{\@secondoftwo}%
\providecommand \translation [1]{[#1]}%
\providecommand \BibitemOpen [0]{}%
\providecommand \bibitemStop [0]{}%
\providecommand \bibitemNoStop [0]{.\EOS\space}%
\providecommand \EOS [0]{\spacefactor3000\relax}%
\providecommand \BibitemShut  [1]{\csname bibitem#1\endcsname}%
\let\auto@bib@innerbib\@empty
\bibitem [{\citenamefont {Tanabashi}\ \emph {et~al.}(2018)\citenamefont
  {Tanabashi} \emph {et~al.}}]{review}%
  \BibitemOpen
  \bibfield  {author} {\bibinfo {author} {\bibfnamefont {M.}~\bibnamefont
  {Tanabashi}} \emph {et~al.} (\bibinfo {collaboration} {Particle Data
  Group}),\ }\href {https://doi.org/10.1103/PhysRevD.98.030001} {\bibfield
  {journal} {\bibinfo  {journal} {Phys. Rev. D}\ }\textbf {\bibinfo {volume}
  {98}},\ \bibinfo {pages} {030001} (\bibinfo {year} {2018})}\BibitemShut
  {NoStop}%
\bibitem [{\citenamefont {Abe}\ \emph {et~al.}(2020)\citenamefont {Abe} \emph
  {et~al.}}]{nature1}%
  \BibitemOpen
  \bibfield  {author} {\bibinfo {author} {\bibfnamefont {K.}~\bibnamefont
  {Abe}} \emph {et~al.} (\bibinfo {collaboration} {T2K Collaboration}),\ }\href
  {https://doi.org/10.1038/s41586-020-2177-0} {\bibfield  {journal} {\bibinfo
  {journal} {Nature}\ }\textbf {\bibinfo {volume} {580}},\ \bibinfo {pages}
  {339} (\bibinfo {year} {2020})}\BibitemShut {NoStop}%
\bibitem [{\citenamefont {Aguilar-Arevalo}\ \emph {et~al.}(2010)\citenamefont
  {Aguilar-Arevalo} \emph {et~al.}}]{AguilarArevalo:2010zc}%
  \BibitemOpen
  \bibfield  {author} {\bibinfo {author} {\bibfnamefont {A.~A.}\ \bibnamefont
  {Aguilar-Arevalo}} \emph {et~al.} (\bibinfo {collaboration} {MiniBooNE
  Collaboration}),\ }\href {https://doi.org/10.1103/PhysRevD.81.092005}
  {\bibfield  {journal} {\bibinfo  {journal} {Phys. Rev. D}\ }\textbf {\bibinfo
  {volume} {81}},\ \bibinfo {pages} {092005} (\bibinfo {year}
  {2010})}\BibitemShut {NoStop}%
\bibitem [{\citenamefont {Aguilar-Arevalo}\ \emph {et~al.}(2013)\citenamefont
  {Aguilar-Arevalo} \emph {et~al.}}]{AguilarArevalo:2013hm}%
  \BibitemOpen
  \bibfield  {author} {\bibinfo {author} {\bibfnamefont {A.~A.}\ \bibnamefont
  {Aguilar-Arevalo}} \emph {et~al.} (\bibinfo {collaboration} {MiniBooNE
  Collaboration}),\ }\href {https://doi.org/10.1103/PhysRevD.88.032001}
  {\bibfield  {journal} {\bibinfo  {journal} {Phys. Rev. D}\ }\textbf {\bibinfo
  {volume} {88}},\ \bibinfo {pages} {032001} (\bibinfo {year}
  {2013})}\BibitemShut {NoStop}%
\bibitem [{\citenamefont {Lyubushkin}\ \emph {et~al.}(2009)\citenamefont
  {Lyubushkin} \emph {et~al.}}]{Lyubushkin:2008pe}%
  \BibitemOpen
  \bibfield  {author} {\bibinfo {author} {\bibfnamefont {V.}~\bibnamefont
  {Lyubushkin}} \emph {et~al.} (\bibinfo {collaboration} {NOMAD
  Collaboration}),\ }\href {https://doi.org/10.1140/epjc/s10052-009-1113-0}
  {\bibfield  {journal} {\bibinfo  {journal} {Eur. Phys. J. C}\ }\textbf
  {\bibinfo {volume} {63}},\ \bibinfo {pages} {355} (\bibinfo {year}
  {2009})}\BibitemShut {NoStop}%
\bibitem [{\citenamefont {Fiorentini}\ \emph {et~al.}(2013)\citenamefont
  {Fiorentini} \emph {et~al.}}]{Minerva1}%
  \BibitemOpen
  \bibfield  {author} {\bibinfo {author} {\bibfnamefont {G.~A.}\ \bibnamefont
  {Fiorentini}} \emph {et~al.} (\bibinfo {collaboration} {MINER$\nu$A
  collaboration}),\ }\href {https://doi.org/10.1103/PhysRevLett.111.022502}
  {\bibfield  {journal} {\bibinfo  {journal} {Phys. Rev. Lett.}\ }\textbf
  {\bibinfo {volume} {111}},\ \bibinfo {pages} {022502} (\bibinfo {year}
  {2013})}\BibitemShut {NoStop}%
\bibitem [{\citenamefont {Fields}\ \emph {et~al.}(2013)\citenamefont {Fields}
  \emph {et~al.}}]{Minerva2}%
  \BibitemOpen
  \bibfield  {author} {\bibinfo {author} {\bibfnamefont {L.}~\bibnamefont
  {Fields}} \emph {et~al.} (\bibinfo {collaboration} {MINER$\nu$A
  collaboration}),\ }\href {https://doi.org/10.1103/PhysRevLett.111.022501}
  {\bibfield  {journal} {\bibinfo  {journal} {Phys. Rev. Lett.}\ }\textbf
  {\bibinfo {volume} {111}},\ \bibinfo {pages} {022501} (\bibinfo {year}
  {2013})}\BibitemShut {NoStop}%
\bibitem [{\citenamefont {Wolcott}\ \emph {et~al.}(2016)\citenamefont {Wolcott}
  \emph {et~al.}}]{Minervanue}%
  \BibitemOpen
  \bibfield  {author} {\bibinfo {author} {\bibfnamefont {J.}~\bibnamefont
  {Wolcott}} \emph {et~al.} (\bibinfo {collaboration} {MINER$\nu$A
  collaboration}),\ }\href {https://doi.org/10.1103/PhysRevLett.116.081802}
  {\bibfield  {journal} {\bibinfo  {journal} {Phys. Rev. Lett.}\ }\textbf
  {\bibinfo {volume} {116}},\ \bibinfo {pages} {081802} (\bibinfo {year}
  {2016})}\BibitemShut {NoStop}%
\bibitem [{\citenamefont {Abe}\ \emph {et~al.}(2013)\citenamefont {Abe} \emph
  {et~al.}}]{T2Kincl}%
  \BibitemOpen
  \bibfield  {author} {\bibinfo {author} {\bibfnamefont {K.}~\bibnamefont
  {Abe}} \emph {et~al.} (\bibinfo {collaboration} {T2K Collaboration}),\ }\href
  {https://doi.org/10.1103/PhysRevD.87.092003} {\bibfield  {journal} {\bibinfo
  {journal} {Phys. Rev. D}\ }\textbf {\bibinfo {volume} {87}},\ \bibinfo
  {pages} {092003} (\bibinfo {year} {2013})}\BibitemShut {NoStop}%
\bibitem [{\citenamefont {Abe}\ \emph {et~al.}(2014)\citenamefont {Abe} \emph
  {et~al.}}]{T2Kinclelectron}%
  \BibitemOpen
  \bibfield  {author} {\bibinfo {author} {\bibfnamefont {K.}~\bibnamefont
  {Abe}} \emph {et~al.} (\bibinfo {collaboration} {T2K Collaboration}),\ }\href
  {https://doi.org/10.1103/PhysRevLett.113.241803} {\bibfield  {journal}
  {\bibinfo  {journal} {Phys. Rev. Lett.}\ }\textbf {\bibinfo {volume} {113}},\
  \bibinfo {pages} {241803} (\bibinfo {year} {2014})}\BibitemShut {NoStop}%
\bibitem [{\citenamefont {Abe}\ \emph {et~al.}(2016)\citenamefont {Abe} \emph
  {et~al.}}]{T2Kcc0pi}%
  \BibitemOpen
  \bibfield  {author} {\bibinfo {author} {\bibfnamefont {K.}~\bibnamefont
  {Abe}} \emph {et~al.} (\bibinfo {collaboration} {T2K Collaboration}),\ }\href
  {https://doi.org/10.1103/PhysRevD.93.112012} {\bibfield  {journal} {\bibinfo
  {journal} {Phys. Rev. D}\ }\textbf {\bibinfo {volume} {93}},\ \bibinfo
  {pages} {112012} (\bibinfo {year} {2016})}\BibitemShut {NoStop}%
\bibitem [{\citenamefont {Acciarri}\ \emph {et~al.}(2014)\citenamefont
  {Acciarri} \emph {et~al.}}]{Argoneutincl2}%
  \BibitemOpen
  \bibfield  {author} {\bibinfo {author} {\bibfnamefont {R.}~\bibnamefont
  {Acciarri}} \emph {et~al.} (\bibinfo {collaboration} {ArgoNeuT
  Collaboration}),\ }\href {https://doi.org/10.1103/PhysRevD.89.112003}
  {\bibfield  {journal} {\bibinfo  {journal} {Phys. Rev. D}\ }\textbf {\bibinfo
  {volume} {89}},\ \bibinfo {pages} {112003} (\bibinfo {year}
  {2014})}\BibitemShut {NoStop}%
\bibitem [{\citenamefont {Abe}\ \emph {et~al.}(2018{\natexlab{a}})\citenamefont
  {Abe} \emph {et~al.}}]{PhysRevD.98.032003}%
  \BibitemOpen
  \bibfield  {author} {\bibinfo {author} {\bibfnamefont {K.}~\bibnamefont
  {Abe}} \emph {et~al.} (\bibinfo {collaboration} {The T2K Collaboration}),\
  }\href {https://doi.org/10.1103/PhysRevD.98.032003} {\bibfield  {journal}
  {\bibinfo  {journal} {Phys. Rev. D}\ }\textbf {\bibinfo {volume} {98}},\
  \bibinfo {pages} {032003} (\bibinfo {year} {2018}{\natexlab{a}})}\BibitemShut
  {NoStop}%
\bibitem [{\citenamefont {Lu}\ and\ \citenamefont
  {others.}(2018)}]{PhysRevLett.121.022504}%
  \BibitemOpen
  \bibfield  {author} {\bibinfo {author} {\bibfnamefont {X.-G.}\ \bibnamefont
  {Lu}}\ and\ \bibinfo {author} {\bibnamefont {others.}} (\bibinfo
  {collaboration} {MINERvA Collaboration}),\ }\href
  {https://doi.org/10.1103/PhysRevLett.121.022504} {\bibfield  {journal}
  {\bibinfo  {journal} {Phys. Rev. Lett.}\ }\textbf {\bibinfo {volume} {121}},\
  \bibinfo {pages} {022504} (\bibinfo {year} {2018})}\BibitemShut {NoStop}%
\bibitem [{\citenamefont {Cai}\ \emph {et~al.}(2020)\citenamefont {Cai} \emph
  {et~al.}}]{PhysRevD.101.092001}%
  \BibitemOpen
  \bibfield  {author} {\bibinfo {author} {\bibfnamefont {T.}~\bibnamefont
  {Cai}} \emph {et~al.} (\bibinfo {collaboration} {The MINER$\nu$A
  Collaboration}),\ }\href {https://doi.org/10.1103/PhysRevD.101.092001}
  {\bibfield  {journal} {\bibinfo  {journal} {Phys. Rev. D}\ }\textbf {\bibinfo
  {volume} {101}},\ \bibinfo {pages} {092001} (\bibinfo {year}
  {2020})}\BibitemShut {NoStop}%
\bibitem [{\citenamefont {Abratenko}\ \emph
  {et~al.}(2020{\natexlab{a}})\citenamefont {Abratenko} \emph
  {et~al.}}]{microbooneA}%
  \BibitemOpen
  \bibfield  {author} {\bibinfo {author} {\bibfnamefont {P.}~\bibnamefont
  {Abratenko}} \emph {et~al.} (\bibinfo {collaboration} {MicroBooNE
  Collaboration}),\ }\href {https://doi.org/10.1103/PhysRevD.102.112013}
  {\bibfield  {journal} {\bibinfo  {journal} {Phys. Rev. D}\ }\textbf {\bibinfo
  {volume} {102}},\ \bibinfo {pages} {112013} (\bibinfo {year}
  {2020}{\natexlab{a}})}\BibitemShut {NoStop}%
\bibitem [{\citenamefont {Abratenko}\ \emph
  {et~al.}(2020{\natexlab{b}})\citenamefont {Abratenko} \emph
  {et~al.}}]{microbooneB}%
  \BibitemOpen
  \bibfield  {author} {\bibinfo {author} {\bibfnamefont {P.}~\bibnamefont
  {Abratenko}} \emph {et~al.} (\bibinfo {collaboration} {MicroBooNE
  Collaboration}),\ }\href {https://doi.org/10.1103/PhysRevLett.125.201803}
  {\bibfield  {journal} {\bibinfo  {journal} {Phys. Rev. Lett.}\ }\textbf
  {\bibinfo {volume} {125}},\ \bibinfo {pages} {201803} (\bibinfo {year}
  {2020}{\natexlab{b}})}\BibitemShut {NoStop}%
\bibitem [{\citenamefont {Simpson}\ \emph {et~al.}(2019)\citenamefont {Simpson}
  \emph {et~al.}}]{Simpson:2019xwo}%
  \BibitemOpen
  \bibfield  {author} {\bibinfo {author} {\bibfnamefont {C.}~\bibnamefont
  {Simpson}} \emph {et~al.} (\bibinfo {collaboration} {Super-Kamiokande}),\
  }\href {https://doi.org/10.3847/1538-4357/ab4883} {\bibfield  {journal}
  {\bibinfo  {journal} {Astrophys. J.}\ }\textbf {\bibinfo {volume} {885}},\
  \bibinfo {pages} {133} (\bibinfo {year} {2019})},\ \Eprint
  {https://arxiv.org/abs/1908.07551} {arXiv:1908.07551 [astro-ph.HE]}
  \BibitemShut {NoStop}%
\bibitem [{\citenamefont {Abe}\ \emph {et~al.}(2018{\natexlab{b}})\citenamefont
  {Abe} \emph {et~al.}}]{protocollaboration2018hyperkamiokande}%
  \BibitemOpen
  \bibfield  {author} {\bibinfo {author} {\bibfnamefont {K.}~\bibnamefont
  {Abe}} \emph {et~al.} (\bibinfo {collaboration} {Hyper-Kamiokande
  Proto-Collaboration}),\ }\href@noop {} {\  (\bibinfo {year}
  {2018}{\natexlab{b}})},\ \Eprint {https://arxiv.org/abs/1805.04163}
  {arXiv:1805.04163 [physics.ins-det]} \BibitemShut {NoStop}%
\bibitem [{\citenamefont {Acciarri}\ \emph {et~al.}(2016)\citenamefont
  {Acciarri} \emph {et~al.}}]{dunecollaboration2016longbaseline}%
  \BibitemOpen
  \bibfield  {author} {\bibinfo {author} {\bibfnamefont {R.}~\bibnamefont
  {Acciarri}} \emph {et~al.} (\bibinfo {collaboration} {DUNE Collaboration}),\
  }\href@noop {} {\  (\bibinfo {year} {2016})},\ \Eprint
  {https://arxiv.org/abs/1512.06148} {arXiv:1512.06148 [physics.ins-det]}
  \BibitemShut {NoStop}%
\bibitem [{\citenamefont {Franco-Patino}\ \emph {et~al.}(2020)\citenamefont
  {Franco-Patino}, \citenamefont {Gonzalez-Rosa}, \citenamefont {Caballero},\
  and\ \citenamefont {Barbaro}}]{PhysRevC.102.064626}%
  \BibitemOpen
  \bibfield  {author} {\bibinfo {author} {\bibfnamefont {J.~M.}\ \bibnamefont
  {Franco-Patino}}, \bibinfo {author} {\bibfnamefont {J.}~\bibnamefont
  {Gonzalez-Rosa}}, \bibinfo {author} {\bibfnamefont {J.~A.}\ \bibnamefont
  {Caballero}},\ and\ \bibinfo {author} {\bibfnamefont {M.~B.}\ \bibnamefont
  {Barbaro}},\ }\href {https://doi.org/10.1103/PhysRevC.102.064626} {\bibfield
  {journal} {\bibinfo  {journal} {Phys. Rev. C}\ }\textbf {\bibinfo {volume}
  {102}},\ \bibinfo {pages} {064626} (\bibinfo {year} {2020})}\BibitemShut
  {NoStop}%
\bibitem [{\citenamefont {Moreno}\ \emph {et~al.}(2014)\citenamefont {Moreno},
  \citenamefont {Donnelly}, \citenamefont {Van~Orden},\ and\ \citenamefont
  {Ford}}]{PhysRevD.90.013014}%
  \BibitemOpen
  \bibfield  {author} {\bibinfo {author} {\bibfnamefont {O.}~\bibnamefont
  {Moreno}}, \bibinfo {author} {\bibfnamefont {T.~W.}\ \bibnamefont
  {Donnelly}}, \bibinfo {author} {\bibfnamefont {J.~W.}\ \bibnamefont
  {Van~Orden}},\ and\ \bibinfo {author} {\bibfnamefont {W.~P.}\ \bibnamefont
  {Ford}},\ }\href {https://doi.org/10.1103/PhysRevD.90.013014} {\bibfield
  {journal} {\bibinfo  {journal} {Phys. Rev. D}\ }\textbf {\bibinfo {volume}
  {90}},\ \bibinfo {pages} {013014} (\bibinfo {year} {2014})}\BibitemShut
  {NoStop}%
\bibitem [{\citenamefont {Van~Orden}\ and\ \citenamefont
  {Donnelly}(2019)}]{PhysRevC.100.044620}%
  \BibitemOpen
  \bibfield  {author} {\bibinfo {author} {\bibfnamefont {J.~W.}\ \bibnamefont
  {Van~Orden}}\ and\ \bibinfo {author} {\bibfnamefont {T.~W.}\ \bibnamefont
  {Donnelly}},\ }\href {https://doi.org/10.1103/PhysRevC.100.044620} {\bibfield
   {journal} {\bibinfo  {journal} {Phys. Rev. C}\ }\textbf {\bibinfo {volume}
  {100}},\ \bibinfo {pages} {044620} (\bibinfo {year} {2019})}\BibitemShut
  {NoStop}%
\bibitem [{\citenamefont {L\"owdin}(1955)}]{PhysRev.97.1474}%
  \BibitemOpen
  \bibfield  {author} {\bibinfo {author} {\bibfnamefont {P.-O.}\ \bibnamefont
  {L\"owdin}},\ }\href {https://doi.org/10.1103/PhysRev.97.1474} {\bibfield
  {journal} {\bibinfo  {journal} {Phys. Rev.}\ }\textbf {\bibinfo {volume}
  {97}},\ \bibinfo {pages} {1474} (\bibinfo {year} {1955})}\BibitemShut
  {NoStop}%
\bibitem [{\citenamefont {Ivanov}\ \emph {et~al.}(2014)\citenamefont {Ivanov},
  \citenamefont {Antonov}, \citenamefont {Caballero}, \citenamefont {Megias},
  \citenamefont {Barbaro}, \citenamefont {Moya~de Guerra},\ and\ \citenamefont
  {Udias}}]{PhysRevC.89.014607}%
  \BibitemOpen
  \bibfield  {author} {\bibinfo {author} {\bibfnamefont {M.~V.}\ \bibnamefont
  {Ivanov}}, \bibinfo {author} {\bibfnamefont {A.~N.}\ \bibnamefont {Antonov}},
  \bibinfo {author} {\bibfnamefont {J.~A.}\ \bibnamefont {Caballero}}, \bibinfo
  {author} {\bibfnamefont {G.~D.}\ \bibnamefont {Megias}}, \bibinfo {author}
  {\bibfnamefont {M.~B.}\ \bibnamefont {Barbaro}}, \bibinfo {author}
  {\bibfnamefont {E.}~\bibnamefont {Moya~de Guerra}},\ and\ \bibinfo {author}
  {\bibfnamefont {J.~M.}\ \bibnamefont {Udias}},\ }\href
  {https://link.aps.org/doi/10.1103/PhysRevC.89.014607} {\bibfield  {journal}
  {\bibinfo  {journal} {Phys. Rev. C}\ }\textbf {\bibinfo {volume} {89}},\
  \bibinfo {pages} {014607} (\bibinfo {year} {2014})}\BibitemShut {NoStop}%
\bibitem [{\citenamefont {Lu}\ \emph {et~al.}(2016)\citenamefont {Lu},
  \citenamefont {Pickering}, \citenamefont {Dolan}, \citenamefont {Barr},
  \citenamefont {Coplowe}, \citenamefont {Uchida}, \citenamefont {Wark},
  \citenamefont {Wascko}, \citenamefont {Weber},\ and\ \citenamefont
  {Yuan}}]{PhysRevC.94.015503}%
  \BibitemOpen
  \bibfield  {author} {\bibinfo {author} {\bibfnamefont {X.-G.}\ \bibnamefont
  {Lu}}, \bibinfo {author} {\bibfnamefont {L.}~\bibnamefont {Pickering}},
  \bibinfo {author} {\bibfnamefont {S.}~\bibnamefont {Dolan}}, \bibinfo
  {author} {\bibfnamefont {G.}~\bibnamefont {Barr}}, \bibinfo {author}
  {\bibfnamefont {D.}~\bibnamefont {Coplowe}}, \bibinfo {author} {\bibfnamefont
  {Y.}~\bibnamefont {Uchida}}, \bibinfo {author} {\bibfnamefont
  {D.}~\bibnamefont {Wark}}, \bibinfo {author} {\bibfnamefont {M.~O.}\
  \bibnamefont {Wascko}}, \bibinfo {author} {\bibfnamefont {A.}~\bibnamefont
  {Weber}},\ and\ \bibinfo {author} {\bibfnamefont {T.}~\bibnamefont {Yuan}},\
  }\href {https://doi.org/10.1103/PhysRevC.94.015503} {\bibfield  {journal}
  {\bibinfo  {journal} {Phys. Rev. C}\ }\textbf {\bibinfo {volume} {94}},\
  \bibinfo {pages} {015503} (\bibinfo {year} {2016})}\BibitemShut {NoStop}%
\bibitem [{\citenamefont {Pickering}(2016)}]{doi:10.7566/JPSCP.12.010032}%
  \BibitemOpen
  \bibfield  {author} {\bibinfo {author} {\bibfnamefont {L.}~\bibnamefont
  {Pickering}},\ }\href {https://doi.org/10.7566/JPSCP.12.010032} {\bibfield
  {journal} {\bibinfo  {journal} {J. Phys. Soc. Jpn. Conf. Proc}\ }\textbf
  {\bibinfo {volume} {12}},\ \bibinfo {pages} {010032} (\bibinfo {year}
  {2016})}\BibitemShut {NoStop}%
\bibitem [{\citenamefont {Gonzalez-Jimenez}\ \emph {et~al.}(2020)\citenamefont
  {Gonzalez-Jimenez}, \citenamefont {Barbaro}, \citenamefont {Caballero},
  \citenamefont {Donnelly}, \citenamefont {Jachowicz}, \citenamefont {Megias},
  \citenamefont {Niewczas}, \citenamefont {Nikolakopoulos},\ and\ \citenamefont
  {Udias}}]{PhysRevC.101.015503}%
  \BibitemOpen
  \bibfield  {author} {\bibinfo {author} {\bibfnamefont {R.}~\bibnamefont
  {Gonzalez-Jimenez}}, \bibinfo {author} {\bibfnamefont {M.~B.}\ \bibnamefont
  {Barbaro}}, \bibinfo {author} {\bibfnamefont {J.~A.}\ \bibnamefont
  {Caballero}}, \bibinfo {author} {\bibfnamefont {T.~W.}\ \bibnamefont
  {Donnelly}}, \bibinfo {author} {\bibfnamefont {N.}~\bibnamefont {Jachowicz}},
  \bibinfo {author} {\bibfnamefont {G.~D.}\ \bibnamefont {Megias}}, \bibinfo
  {author} {\bibfnamefont {K.}~\bibnamefont {Niewczas}}, \bibinfo {author}
  {\bibfnamefont {A.}~\bibnamefont {Nikolakopoulos}},\ and\ \bibinfo {author}
  {\bibfnamefont {J.~M.}\ \bibnamefont {Udias}},\ }\href
  {https://doi.org/10.1103/PhysRevC.101.015503} {\bibfield  {journal} {\bibinfo
   {journal} {Phys. Rev. C}\ }\textbf {\bibinfo {volume} {101}},\ \bibinfo
  {pages} {015503} (\bibinfo {year} {2020})}\BibitemShut {NoStop}%
\bibitem [{\citenamefont {Megias}\ \emph {et~al.}(2018)\citenamefont {Megias},
  \citenamefont {Barbaro}, \citenamefont {Caballero}, \citenamefont {Amaro},
  \citenamefont {Donnelly}, \citenamefont {Simo},\ and\ \citenamefont
  {Orden}}]{Megias_2018}%
  \BibitemOpen
  \bibfield  {author} {\bibinfo {author} {\bibfnamefont {G.~D.}\ \bibnamefont
  {Megias}}, \bibinfo {author} {\bibfnamefont {M.~B.}\ \bibnamefont {Barbaro}},
  \bibinfo {author} {\bibfnamefont {J.~A.}\ \bibnamefont {Caballero}}, \bibinfo
  {author} {\bibfnamefont {J.~E.}\ \bibnamefont {Amaro}}, \bibinfo {author}
  {\bibfnamefont {T.~W.}\ \bibnamefont {Donnelly}}, \bibinfo {author}
  {\bibfnamefont {I.~R.}\ \bibnamefont {Simo}},\ and\ \bibinfo {author}
  {\bibfnamefont {J.~W.~V.}\ \bibnamefont {Orden}},\ }\href
  {https://doi.org/10.1088/1361-6471/aaf3ae} {\bibfield  {journal} {\bibinfo
  {journal} {Journal of Physics G: Nuclear and Particle Physics}\ }\textbf
  {\bibinfo {volume} {46}},\ \bibinfo {pages} {015104} (\bibinfo {year}
  {2018})}\BibitemShut {NoStop}%
\end{thebibliography}%

\end{document}